\documentclass[twocolumn, twocolappendix]{aastex631}
\usepackage{enumitem}
\usepackage{xspace}
\usepackage{amsmath}
\usepackage{float}

\newcommand{\sw}[1]{\texttt{#1}}

\shorttitle{Probing baryonic feedback and cosmology with 3$\times$2-point statistic of FRBs and galaxies}
\shortauthors{Sharma et al.}

\begin{document}

\title{Probing baryonic feedback and cosmology with 3$\times$2-point statistic of FRBs and galaxies}

\correspondingauthor{Kritti Sharma}
\email{kritti@caltech.edu}

\author[0000-0002-4477-3625]{Kritti Sharma}
\affiliation{Cahill Center for Astronomy and Astrophysics, MC 249-17 California Institute of Technology, Pasadena CA 91125, USA.}

\author[0000-0001-8356-2014]{Elisabeth Krause}
\affiliation{Department of Astronomy/Steward Observatory, University of Arizona, 933 North Cherry Avenue, Tucson, AZ 85721, USA}
\affiliation{Department of Physics, University of Arizona, 1118 E Fourth Street, Tucson, AZ 85721,
USA}

\author[0000-0002-7252-5485]{Vikram Ravi}
\affiliation{Cahill Center for Astronomy and Astrophysics, MC 249-17 California Institute of Technology, Pasadena CA 91125, USA.}
\affiliation{Owens Valley Radio Observatory, California Institute of Technology, Big Pine CA 93513, USA.}

\author[0000-0001-5404-8753]{Robert Reischke}
\affiliation{Argelander-Institut für Astronomie, Universität Bonn, Auf dem Hügel 71, D-53121 Bonn, Germany}

\author[0000-0002-7587-6352]{Liam Connor}
\affiliation{Center for Astrophysics | Harvard $\&$ Smithsonian, Cambridge, MA 02138-1516, USA.}

\author[0000-0003-3714-2574]{Pranjal R. S.}
\affiliation{Department of Astronomy/Steward Observatory, University of Arizona, 933 North Cherry Avenue, Tucson, AZ 85721, USA}

\author[0000-0003-3312-909X]{Dhayaa Anbajagane}
\affiliation{Department of Astronomy and Astrophysics, University of Chicago, Chicago, IL 60637, USA}
\affiliation{Kavli Institute for Cosmological Physics, University of Chicago, Chicago, IL 60637, USA}

\begin{abstract}

The impact of galaxy formation processes on the matter power spectrum is uncertain and may bias cosmological parameters inferred by large-scale structure surveys. Fast Radio Bursts (FRBs), through their dispersion measures (DMs) encoding the integrated column density of baryons, offer a unique window into the distribution of gas on scales $k \lesssim 10~h$\,Mpc$^{-1}$. In this work, we investigate the constraining power of a 3×2-point correlation statistic of FRB DMs and galaxy number density fields. We present the formalism for correlation analyses, derive the associated covariance matrices and forecast signal-to-noise ratios and Fisher parameter constraints. Assuming host galaxy DM variance of 90~pc\,cm$^{-3}$, for $10^4$ ($10^5$) FRBs across 35\% of the sky, the angular DM power spectrum is noise dominated at multipoles $\ell \gtrsim 20$ ($\ell \gtrsim 100$), which implies that the analysis can be conducted using arcmin-scale localizations, where the redshift distribution of the FRB population can be modeled through the FRB luminosity function or FRB position cross-correlations with galaxies. We show that while $10^4$ ($10^5$) FRB DM correlations can constrain cosmological parameters at $\sim$40-70\% (30-40\%) level, this is a factor of $\sim$2-3 ($1.5$-2) weaker than the precision attainable with galaxy clustering alone due to shot noise from the limited FRB number density $\sim$0.5 ($\sim$5)\,deg$^{-2}$, variance of the field and host DMs. On the contrary, feedback-sensitive non-linear scales are not currently accessible in galaxy surveys. We demonstrate that combining FRB DM auto-correlations, FRB DM--galaxy cross-correlations, and galaxy auto-correlations in a 3×2-point analysis breaks feedback-cosmology degeneracies and significantly improves constraints, yielding $\sim$10-18\% (7-13\%) precision on cosmological parameters and $\sim$3\% (2\%) constraints on feedback using $10^4$ ($10^5$) FRBs. This work positions the 3×2-point statistic of FRB DMs and galaxies as a promising multi-probe strategy, bridging the gap between constraining astrophysical feedback models and precise measurement of cosmological parameters.

\end{abstract}

\section{Introduction} \label{sec:introduction}

Precision cosmology stands at a crossroads: the next generation of large-scale structure surveys promise percent-level constraints on cosmological parameters~\citep{2011arXiv1110.3193L, 2013arXiv1305.5425S, 2018arXiv181200514F, 2019ApJ...873..111I, 2020SPIE11443E..0IC, 2024AJ....168...58D}, but further progress is hampered by our limited understanding of small-scale baryonic physics~\citep{2012JCAP...04..034H, 2022PhRvD.105b3514A, 2023MNRAS.525.5554P, 2024JCAP...07..037T}. Of particular concern is the impact of energetic feedback from galaxy formation processes on the matter distribution at small scales~\citep{2019OJAp....2E...4C}. Accurately modeling these effects is critical, not just for controlling systematics in clustering and lensing measurements, but also for potentially uncovering new physics through deviations from the standard $\Lambda$ cold dark matter ($\Lambda$CDM) cosmological model~\citep{2025arXiv250401669D}.

Large-scale structure surveys, including the Dark Energy Spectroscopic Instrument~\citep[DESI;][]{2022AJ....164..207D}, Vera Rubin Observatory~\citep[][]{2019ApJ...873..111I}, Euclid~\citep{2011arXiv1110.3193L, 2022A&A...662A.112E}, Nancy Grace Roman Space Telescope~\citep[\textit{Roman};][]{2013arXiv1305.5425S, 2018arXiv181200514F} and SPHEREx~\citep{2020SPIE11443E..0IC}, are designed to evaluate the plausibility of deviations from the standard $\Lambda$CDM cosmological model. Combining independent probes of gas distribution with large-scale structure surveys can constrain baryonic feedback, together with cosmology. For example, joint analyses of galaxies and cosmic shear with X-ray observations~\citep[e.g.][]{2022MNRAS.514.3802S, 2024A&A...690A.267Z, 2024PhRvL.133e1001F, 2024A&A...689A.298G, 2024MNRAS.528.4379G, 2024eas..conf.1906L, 2024eas..conf.1910L, 2025ApJ...983....8L,2025arXiv250707991K}, the kinematic Sunyaev-Zel’dovich~\citep[kSZ; e.g.][]{2021PhRvD.104d3502C, 2022MNRAS.514.3802S, 2024MNRAS.534..655B, 2024arXiv240707152H,2025arXiv250319870R,2025arXiv250714136H} effect and the thermal Sunyaev-Zel’dovich~\citep[tSZ; e.g.][]{2022A&A...660A..27T, 2023MNRAS.525.1779P, 2024OJAp....7E..13B,2025arXiv250607432P,2025arXiv250704476D} effect have helped constrain the impact of baryonic feedback at small scales. 

Fast Radio Bursts~\citep[FRBs;][]{2022A&ARv..30....2P} have recently emerged as a powerful complementary probe of the baryonic universe~\citep{2007Sci...318..777L, 2013Sci...341...53T, 2014ApJ...780L..33M}. Their dispersion measures (DMs), which encode the integrated free electron column density along the line of sight, offer a unique window into the distribution of ionized baryons in the cosmic web~\citep{2014ApJ...780L..33M}. The variance in DM across different sightlines for FRB sources at the same redshift (1-point statistic) is sensitive to the effects of feedback processes on the matter power spectrum~\citep{2025arXiv250418745S,2025arXiv250717742R}. Early efforts have used the DM--redshift relation as a one-point statistic to infer cosmological parameters and feedback strength~\citep{2020Natur.581..391M, 2022MNRAS.511..662H, 2022MNRAS.509.4775J, 2022MNRAS.516.4862J, 2024ApJ...965...57B, 2024arXiv240200505K, 2024arXiv240916952C,2024arXiv241117682R}. However, the approaches measuring probability density function $p(\mathrm{DM}, z)$ are susceptible to observational biases and selection effects, such as telescope detection efficiency and FRB population characteristics~\citep{2022MNRAS.509.4775J, 2022MNRAS.516.4862J}\footnote{The impact of these selection effects on inference using $p(\mathrm{DM}|z)$~\citep{2020Natur.581..391M, 2024arXiv240916952C} is unclear.}.

The impact of instrument selection and propagation effects on 1-point statistic~\citep{2022MNRAS.509.4775J, 2022MNRAS.516.4862J} and angular FRB position cross-correlations with galaxy positions~\citep{2020PhRvD.102b3528R, 2021ApJ...922...42R} has motivated growing interest in measuring the angular cross-correlation of FRB DMs with galaxy positions~\citep{2017PhRvD..95h3012S}, which is conceptually similar to galaxy-galaxy lensing\footnote{Galaxy-galaxy lensing is the cross-correlation between the positions of foreground (lens) galaxies and the shear field measured from background (source) galaxies, which measures the matter distribution around lens galaxies~\citep{2023MNRAS.524.6071Y}.} and enhances sensitivity to lower mass halos, as compared to the 1-point statistic. These cross-correlations are largely robust to host galaxy properties, as well as DM dependent and scattering selection effects~\citep{2025arXiv250603258Q}, since FRBs are a background source of light and the cross-correlation signal is picked up from the electron distribution at the location of the foreground galaxies. The recent detection of galaxy--DM cross-correlation signal by CHIME/FRB has established the feasibility of this approach~\citep{2025arXiv250608932W}.

Given these developments, it is timely to assess the constraining power of galaxy--DM cross-correlations on baryonic feedback and cosmology. With several high-cadence, wide-area FRB surveys now underway or imminent, such as the Deep Synoptic Array~\citep[DSA-2000;][]{2019BAAS...51g.255H}, the Square Kilometer Array~\citep[SKA;][]{2004NewAR..48..979C} and the Canadian Hydrogen Observatory and Radio-Transient Detector~\citep[CHORD;][]{2019clrp.2020...28V}, we are entering a regime where statistical measurements of FRB DM clustering will be informative. At the same time, it is crucial to understand how these measurements complement existing galaxy clustering observables, and what combinations of statistics yield optimal constraints on both cosmology and baryonic physics.

In this work, we review the auto- and cross-correlation formalism for FRB DMs and galaxy positions, and evaluate their sensitivity to baryonic feedback and cosmological parameters. We perform Fisher information matrix forecasts, discuss the degeneracy structure of various observables, and highlight the gains achievable through joint 3$\times$2-point statistic. This article is structured as follows. In \S\ref{sec:formalism}, we present the theoretical framework for the angular correlation statistics of FRB DMs and galaxies. In \S\ref{sec:simulation_specifications}, we outline our simulation setup and survey specifications. In \S\ref{sec:results_discussion}, we present signal-to-noise ratio and Fisher forecasts for individual and joint observables. In \S\ref{sec:discussion}, we discuss how such an analysis may be carried out in practice, connecting it to lessons from cosmic shear analyses. We conclude in \S\ref{sec:conclusion} with a discussion of implications and prospects for future FRB surveys. Throughout the text, we assume \citet{2020A&A...641A...6P} TT,TE,EE+lowE+lensing cosmology.

\section{Statistical Formalism for Correlation Analyses} \label{sec:formalism}

In this section, we describe the theoretical framework underlying the auto- and cross-correlation statistics for FRB DMs and galaxies. We begin by describing the FRB DM formalism and define the DM perturbation $\mathcal{D}(\hat{x}, z)$, along a sightline $\hat{x}$ out to redshift $z$, in Section~\ref{subsec:FRB_DM}. Next, in Section~\ref{subsec:observables}, we define the two key observables of interest: the angular DM perturbations averaged over the sample of observed FRBs $\mathcal{D}_\mathrm{2D}(\hat{x})$ and the projected galaxy over-density field $\delta_\mathrm{g,2D}(\hat{x})$. In Section~\ref{subsec:one_point_two_point_statistics}, we define the three statistics: (i) the angular DM auto-power spectrum $C^{\mathcal{DD}}$, (ii) the angular galaxy auto-power spectrum $C^{\mathrm{gg}}$, and (iii) the angular DM-galaxy cross-power spectrum $C^{\mathrm{g}\mathcal{D}}$. We delineate the covariance matrices associated with analyses of each statistic: (i) $C^{\mathcal{DD}}$, (ii) $C^{\mathrm{gg}}$, (iii) $C^{\mathrm{g}\mathcal{D}}$, (iv) $(C^{\mathcal{DD}}, C^{\mathrm{g}\mathcal{D}})$ joint analysis, and (v) $(C^{\mathcal{DD}}, C^{\mathrm{gg}}, C^{\mathrm{g}\mathcal{D}})$ joint analysis in Section~\ref{subsec:covariance_matrices}. We refer to the $(C^{\mathcal{DD}}, C^{\mathrm{gg}}, C^{\mathrm{g}\mathcal{D}})$ joint analysis as the 3$\times$2-point statistic. Finally, in Section~\ref{subsec:signal_to_noise_ratio_and_fisher_matrix_formalism}, we outline the signal-to-noise ratio and Fisher information matrix formalism.

\subsection{FRB Dispersion Measure} \label{subsec:FRB_DM}

As radio waves travel from a source at redshift $z_\mathrm{s}$ along direction $\hat{x}$ to an observer, they get progressively dispersed by electrons in the Milky Way, intergalactic medium and the circumgalactic medium of intervening halos and the host galaxy: 
\begin{equation}
    \mathrm{DM}_\mathrm{FRB} (\hat{x},z_\mathrm{s}) = \mathrm{DM}_\mathrm{MW}(\hat{x}) + \mathrm{DM}_\mathrm{cosmic}(\hat{x}, z_\mathrm{s}) + \frac{\mathrm{DM}_\mathrm{host}}{1+z_\mathrm{s}}.
\end{equation}
The Milky Way component ($\mathrm{DM}_\mathrm{MW}$) can be removed from the observed DMs with $\lesssim 10$~pc\,cm$^{-3}$ uncertainty for sightlines not intersecting the galactic plane~\citep{2002astro.ph..7156C, 2003astro.ph..1598C, 2019MNRAS.485..648P, 2023arXiv230101000R}. The last two terms represent the extragalactic DM contribution ($\mathrm{DM}_\mathrm{exgal}$). The host galaxy component (DM$_\mathrm{host}$) includes the contribution of the circumburst medium, interstellar medium, and halo of the host galaxy. The contribution to DM from the intergalactic medium and the circumgalactic medium of intervening halos -- collectively denoted as $\mathrm{DM}_\mathrm{cosmic}$ -- is expressed as an integral over the electron number density $n_\mathrm{e}$ along the line-of-sight,
\begin{equation}
\mathrm{DM}_\mathrm{cosmic} (\hat{x}, z_\mathrm{s}) = \int\limits_0^{\mathrm{s}(z_\mathrm{s})} \frac{n_\mathrm{e} (\hat{x},z)}{(1+z)} \mathrm{d}s,
\label{eqn:DM_definition}
\end{equation}
where $\mathrm{d}s = \mathrm{d}\chi/(1+z)$ is the differential of the proper distance, $\mathrm{d}\chi = c~\mathrm{d}z/H(z)$ is the differential of the comoving distance, $H(z) = H_0 \sqrt{\Omega_\mathrm{m0}(1+z)^3 + \Omega_{\Lambda 0}}$ is the Hubble parameter, $H_0$ is the Hubble constant, $\Omega_\mathrm{m0}$ and $\Omega_{\Lambda 0}$ are the present epoch matter and dark energy densities, respectively. The electron number density at redshift $z$ along direction $\hat{x}$ can be decomposed into a mean component and a fluctuation, 
\begin{equation}
    n_\mathrm{e} (\hat{x}, z) = \bar{n}_\mathrm{e} (z) (1+\delta_\mathrm{e}(\hat{x}, z)),
\end{equation}
where $\delta_\mathrm{e}$ is the dimensionless electron density contrast and the mean electron density is given by,
\begin{equation}
    \bar{n}_\mathrm{e}(z) = \frac{f_\mathrm{d}(z) \rho_\mathrm{b}(z)\chi_\mathrm{e}}{m_\mathrm{p}},
\end{equation} 
where $f_\mathrm{d}(z)$ is the fraction of baryons in the diffuse ionized gas, $\chi_\mathrm{e} = Y_\mathrm{H} + Y_\mathrm{He}/2 \approx 1-Y_\mathrm{He}/2$ is the electron fraction computed using the primordial abundance of helium~\citep{2020A&A...641A...6P}, $m_\mathrm{p}$ is the proton mass, $\rho_\mathrm{b}(z) = \Omega_\mathrm{b0} \rho_\mathrm{c}(1+z)^3$ is the baryon mass density, $\Omega_\mathrm{b0}$ is the present epoch baryon density and $\rho_\mathrm{c} = 3H_0^2/8\pi G$ is the critical density of the Universe. Substituting these into Equation~\ref{eqn:DM_definition}, 
\begin{equation}
\begin{aligned}
    & \mathrm{DM}_\mathrm{cosmic} (\hat{x}, z_\mathrm{s}) \\
    & = \int\limits_0^{z_\mathrm{s}} \frac{3c \chi_\mathrm{e} \Omega_\mathrm{b0} H_0}{8 \pi G m_\mathrm{p}} \frac{f_\mathrm{d}(z) (1+z) (1+\delta_\mathrm{e}(\hat{x}, z))}{\sqrt{\Omega_\mathrm{m0}(1+z)^3 + \Omega_{\Lambda 0}}} \mathrm{d}z \\
    & = \int\limits_0^{z_\mathrm{s}} W_\mathrm{DM}(z) (1+\delta_\mathrm{e}(\hat{x}, z)) \mathrm{d}z \\
    & = \langle \mathrm{DM}_\mathrm{cosmic}(z_{\mathrm{s}}) \rangle + \mathcal{D}(\hat{x}, z_{\mathrm{s}}),
\end{aligned}
\label{eqn:meanDM}
\end{equation}
where $W_\mathrm{DM}(z)$ is the dispersion measure weighting function that explicitly depends on diffuse baryon fraction and cosmological constants, $\langle \mathrm{DM}_\mathrm{cosmic}(z_{\mathrm{s}}) \rangle$ is the mean $\mathrm{DM}_\mathrm{cosmic}$ out to source redshift, and $\mathcal{D}(\hat{x}, z_{\mathrm{s}})$ is the DM perturbation, statistics of which are governed by the degree of clustering of baryons~\citep{2014ApJ...780L..33M, 2023MNRAS.524.2237R, 2025arXiv250418745S}.

\subsection{FRB and Galaxy Observables} \label{subsec:observables}

Given the three-dimensional distribution of objects (FRBs/galaxies) $n_\mathrm{o}(\hat{x}, z)$, the angular object over-density field $\delta_\mathrm{o,2D}(\hat{x})$ can be written as
\begin{equation}
    \delta_\mathrm{o,2D}(\hat{x}) = \frac{n_\mathrm{o,2D}(\hat{x})}{\bar{n}_\mathrm{o,2D}} - 1 = \int_0^\infty \mathrm{d}z~W_\mathrm{o}(z) \delta_\mathrm{o}(\hat{x}, z),
    \label{eqn:angular_overdensity_field}
\end{equation}
where $n_\mathrm{o,2D}$ is the angular object number density, $\bar{n}_\mathrm{o,2D}$ is the average angular object number density, $\delta_\mathrm{o}$ is the dimensionless object number density contrast and $W_\mathrm{o}(z)$ is the object weight function, 
\begin{equation}
    W_\mathrm{o}(z) = \frac{1}{\bar{n}_\mathrm{o,2D}} \frac{\chi^2(z)}{H(z)(1+z)} \bar{n}_\mathrm{o}(z),
    \label{eqn:source_wt_fxn}
\end{equation}
where $\bar{n}_\mathrm{o}(z)$ is the average comoving number density of objects at redshift $z$. Therefore, given the redshift distribution of FRB sources $n_\mathrm{f}(z)$ and galaxies $n_\mathrm{g}(z)$, their angular source over-density fields, $\delta_\mathrm{f,2D}(\hat{x})$ and $\delta_\mathrm{g,2D}(\hat{x})$, can be written in terms of the weighting functions, $W_\mathrm{f}(z)$ and $W_\mathrm{g}(z)$, respectively, defined by Equation~\ref{eqn:angular_overdensity_field} and \ref{eqn:source_wt_fxn}.

Given the observed three-dimensional DM distribution DM$_\mathrm{cosmic}(\hat{x}, z)$, the angular DM perturbation (or, alternatively, the projected electron over-density) field, averaged over a sample of FRBs with mean projected source number density $\bar{n}_\mathrm{f,2D}$, can be written as:
\begin{equation}
\begin{aligned}
    \mathcal{D}_\mathrm{2D}(\hat{x}) & = \frac{1}{\bar{n}_\mathrm{f,2D}} \int_0^{z_\mathrm{H}} \mathrm{d}z~\mathcal{D}(\hat{x}, z) \frac{\chi^2(z)}{H(z)(1+z)}\bar{n}_\mathrm{f}(z) \\
    & = \int_0^{z_\mathrm{H}} \mathrm{d}z~\int_0^z \mathrm{d}z^\prime W_\mathrm{DM}(z^\prime) \delta_\mathrm{e}(\hat{x}, z^\prime) W_\mathrm{f}(z),
    \label{eqn:DM_2D}
\end{aligned}
\end{equation}
where in the second equality we used the definition of DM perturbation $\mathcal{D}(\hat{x}, z)$, and FRB source weight function $W_\mathrm{f}(z)$, from Equation~\ref{eqn:meanDM} and \ref{eqn:source_wt_fxn}, respectively. The integration limit $z_\mathrm{H}$ is the maximum redshift in FRB sample (the horizon). Rearranging the integration limits yields
\begin{equation}
\begin{aligned}
    \mathcal{D}_\mathrm{2D}(\hat{x}) & = \int_0^{z_\mathrm{H}} \mathrm{d}z~\frac{\mathrm{d}\langle \mathrm{DM}_\mathrm{cosmic}(z) \rangle}{\mathrm{d}z} \delta_\mathrm{e}(\hat{x}, z)~\int_z^{z_\mathrm{H}} \mathrm{d}z^\prime W_\mathrm{f}(z^\prime) \\
    & = \int_0^{z_\mathrm{H}} \mathrm{d}z W_{\mathcal{D}}(z) \delta_\mathrm{e}(\hat{x}, z),
    \label{eqn:mean_projected_DM_perturbations}
\end{aligned}
\end{equation}
where the DM perturbation weighting function,
\begin{equation}
\begin{aligned}
W_{\mathcal{D}}(z) = \frac{\mathrm{d}\langle \mathrm{DM}_\mathrm{cosmic}(z) \rangle}{\mathrm{d}z} ~\int_z^{z_\mathrm{H}} \mathrm{d}z^\prime W_\mathrm{f}(z^\prime).
\label{eqn:DM_perturbation_wt_fxn}
\end{aligned}
\end{equation}
Intuitively, this weighting function implies that an electron over-density at redshift $z$ contributes to the DM perturbation observed for any FRB beyond redshift $z$.

\subsection{One-Point and Two-Point Statistics} \label{subsec:one_point_two_point_statistics}

Having defined all observables, we now define the auto- and cross-correlation metrics. We define a general formalism for computing the angular cross-power spectrum between two projected fields, $A(\hat{x})$ and $B(\hat{x})$, which are obtained by line-of-sight integrals of their respective three-dimensional fluctuations, $\delta_A(\hat{x}, z)$ and $\delta_B(\hat{x}, z)$, weighted by functions $W_A(z)$ and $W_B(z)$, respectively (see Equation~\ref{eqn:angular_overdensity_field} and \ref{eqn:mean_projected_DM_perturbations} for reference). The angular cross-power spectrum $C^{AB}(\ell)$ is defined through the spherical harmonic coefficients $a_{\ell m}^A$ and $a_{\ell m}^B$ as
\begin{equation}
    \langle a_{\ell m}^{A} a_{\ell' m'}^{B\ast} \rangle = \delta^\mathrm{K}_{\ell\ell'} \delta^\mathrm{K}_{mm'} C^{AB}(\ell),
\label{eqn:general_power_spectrum_definition}
\end{equation}
where $\delta^\mathrm{K}_{ij}$ is the Kronecker delta function. The spherical harmonic coefficients are computed by expanding the density contrast in Fourier space, taking the spherical harmonic transform and expanding the exponential into plane waves, thus yielding
\begin{equation}
    \label{eqn:alm_DM}
\begin{split}
    a_{\ell m}^A = &\; 4\pi\mathrm{i}^\ell\int_0^\infty \mathrm{d}z\, W_{A}(z) \int_0^\infty \frac{\mathrm{d}^3k}{(2\pi)^3} \tilde{\delta}_{A}(k, z) \\ &\times  j_\ell(k\chi) Y_{\ell m}^\ast (\hat{k}),    
\end{split}
\end{equation}
where $\tilde{\delta}_A$ denotes the Fourier transform of the density contrast field, $k$ denotes the wavenumber and $j_\ell$ denotes the spherical Bessel function of zeroth order. Plugging this back into Equation~\ref{eqn:general_power_spectrum_definition} and using the ortho-normality of spherical harmonics,
\begin{equation}
\begin{aligned}
    C^{AB}(\ell) = & \int\limits_0^\infty k^2 \mathrm{d}k \int\limits_0^{\chi_\mathrm{H}} \mathrm{d}\chi W_{A}(\chi) j_\ell(k\chi) \\
    & \int\limits_0^{\chi_\mathrm{H}} \mathrm{d}\chi^\prime W_{B}(\chi^\prime) j_\ell(k\chi^\prime) {P_{AB} (k, z(\chi), z(\chi^\prime))}, 
\end{aligned}
\end{equation}
where $\chi_\mathrm{H}$ is the comoving distance at the maximum redshift in sample, $P_{AB}(k,z)$ is the cross-power spectrum of the fields $A$ and $B$, defined as
\begin{equation}
    \langle \tilde{\delta}_A(k, z) \tilde{\delta}_B(k^\prime, z) \rangle = (2\pi)^3 \delta_\mathrm{D}(k+k^\prime) P_{AB}(k,z),
\end{equation}
and $\delta_\mathrm{D}$ is the Dirac delta function. For sufficiently large multipole $\ell$, the spherical Bessel functions are highly peaked around $k\chi = \ell$ and therefore, $j_\ell(k\chi) \approx \delta_\mathrm{D}(k-\ell/\chi)/k\chi$. Using this Limber approximation,
\begin{equation}
\begin{aligned}
    & C^{AB}(\ell) \\
    & = \int_0^{\chi_\mathrm{H}} \mathrm{d}\chi\, \frac{W_A(\chi) W_B(\chi)}{\chi^2} P_{AB}\left(k = \frac{\ell + 1/2}{\chi}, z(\chi)\right).
\end{aligned}
    \label{eqn:general_limber}
\end{equation}
When $A = B$, this expression corresponds to the angular auto-power spectrum of field $A$. This generalized formalism encompasses all three angular statistics considered in this work. In what follows, we instantiate this general expression for the relevant field pairs.

\subsubsection{DM Auto-Correlation} \label{subsubsec:DM_autocorrelation}

Using the general formalism introduced above, we set $A=B=\mathcal{D}$ to obtain the angular DM auto-power spectrum $C^{\mathcal{DD}}(\ell)$. The DM perturbation weighting function $W_{\mathcal{D}}(z)$ is defined through the projected DM field $\mathcal{D}(\hat{x})$, and the three-dimensional field fluctuations correspond to the electron density contrast, $\delta_\mathrm{e}(\hat{x}, z)$ (see Equation~\ref{eqn:mean_projected_DM_perturbations}). The corresponding three-dimensional power spectrum is the electron auto-power spectrum $P_\mathrm{ee}(k, z)$. Under the Limber approximation,
\begin{equation}
    C^{\mathcal{DD}}(\ell) = \int_0^{\chi_\mathrm{H}} \mathrm{d}\chi\, \frac{W_{\mathcal{D}}^2(\chi)}{\chi^2} P_\mathrm{ee}\left(k = \frac{\ell + 1/2}{\chi}, z(\chi)\right).
    \label{eqn:DMDM}
\end{equation}
The angular power spectrum serves as the basis to describe the statistics of observed DMs. For reference, the sightline-to-sightline variance in DM is given by the correlation at zero angular separation of FRB sightlines~\citep{2023MNRAS.524.2237R}. This statistic is sensitive to baryonic feedback and cosmology through $P_\mathrm{ee}(k)$ and $W_\mathrm{DM}(\chi)$ (see Equation~\ref{eqn:meanDM} and \ref{eqn:mean_projected_DM_perturbations}).

\subsubsection{Galaxies Auto-Correlation} \label{subsubsec:galaxies_autocorrelation}

Setting $A = B = \mathrm{g}$, the angular galaxy auto-power spectrum $C^\mathrm{gg}(\ell)$ probes the clustering of galaxies. The weighting function $W_{\mathrm{g}}(z)$ is determined by the redshift distribution of galaxies (see Equation~\ref{eqn:source_wt_fxn}), and the 3D field is the galaxy overdensity $\delta_\mathrm{g}(\hat{x}, z)$, with corresponding power spectrum $P_\mathrm{gg}(k, z)$. The Limber-approximated angular power spectrum is given by
\begin{equation}
    C^{\mathrm{gg}}(\ell) = \int_0^{\chi_\mathrm{H}} \mathrm{d}\chi\, \frac{W_{\mathrm{g}}^2(\chi)}{\chi^2} P_\mathrm{gg}\left(k = \frac{\ell + 1/2}{\chi}, z(\chi)\right).
\end{equation}
Since galaxies are biased tracers, in addition to underlying cosmology, modeling their power spectrum requires a few more ingredients and in this work, we take the halo occupation distribution (HOD) approach (see \citet{2002ApJ...575..587B, 2005ApJ...633..791Z, 2007ApJ...667..760Z} for a review of HOD approach and see \citet{2023OJAp....6E..39A} for a review of discrete-tracer halo model approach).

\subsubsection{DM-Galaxies Cross-Correlation} \label{subsubsec:DM_DM_autocorrelation}

To compute the cross-correlation between the FRB DM field and the galaxy field, $C^{\mathrm{g}\mathcal{D}}(\ell)$, we set $A = \mathrm{g}$ and $B = \mathcal{D}$. The relevant 3D cross-spectrum is $P_\mathrm{ge}(k, z)$, the galaxy-electron cross-power spectrum. Under the Limber approximation, the angular cross-power spectrum is
\begin{equation}
    C^{\mathrm{g}\mathcal{D}}(\ell) = \int\limits_0^{\chi_\mathrm{H}} \mathrm{d}\chi 
    \frac{W_{\mathcal{D}}(\chi) W_{\mathrm{g}}(\chi)} 
    {\chi^2} P_\mathrm{ge}\left(k = \frac{\ell+1/2}{\chi}, z(\chi)\right),
    \label{eqn:gDM}
\end{equation}
This cross-correlation probes the co-spatial distribution of galaxies and free electrons, and encodes complementary information to the auto-power spectra.

\subsection{Covariance Matrices} \label{subsec:covariance_matrices}

The Gaussian covariance~\citep[using the definition of covariance and Wick’s theorem for Gaussian fields;][]{2017MNRAS.470.2100K} for two multi-probe power spectra can be written as
\begin{equation}
\begin{aligned}
    \mathrm{Cov}&[C^{AB}(\ell_i),C^{CD}(\ell_j)] =  \frac{\delta^\mathrm{K}_{ij}}{(2\ell_i+1)\Delta\ell f_\mathrm{sky}} \\ \times & [ \hat{C}^{AC}(\ell_i)\hat{C}^{BD} (\ell_j) + \hat{C}^{AD}(\ell_i)\hat{C}^{BC} (\ell_j)]
\;, 
\label{eqn:generalized_covariance}
\end{aligned}
\end{equation}
where $\Delta \ell$ is the width used for binning the power spectra and $f_\mathrm{sky}$ is the fraction of observed sky. With $\hat{C}^{AB}$, we denote an observed power spectrum containing a noise contribution so that:
\begin{equation}
    \hat{C}^{AB}(\ell) = {C}^{AB}(\ell) + \delta^\mathrm{K}_{AB}N_A\;,
\end{equation}
where $N_A$ is the noise contribution for the respective field. For the observed galaxy auto-correlation this is given by
\begin{equation}
    N_\mathrm{g} = \frac{1}{\bar{n}_{\mathrm{g,2D}}},
\end{equation}
where the shot noise depends only on the mean galaxy angular number density. The noise contribution to the observed DM auto-correlation is the sum of the variance of the DM field, $\mathcal{D}$ and the host contribution:
\begin{equation}
\label{eq:noise_level_dm}
    N_\mathcal{D} = \frac{\sigma_{\mathcal{D}}^2}{\bar{n}_{\mathrm{f,2D}}} + \frac{\sigma^2_\mathrm{host}}{(1+\bar{z}_\mathrm{f})^2\bar{n}_{\mathrm{f,2D}}}.
\end{equation}
Here, $\sigma^2_\mathrm{host}$ is the variance in rest-frame host DM contribution, $\bar{z}_\mathrm{f}$ is the mean redshift of the observed FRB sample that transforms the rest-frame host DM variance into the observed DM, $\bar{n}_{\mathrm{f,2D}}$ is the mean number density of FRBs and  the DM field variance is
\begin{equation}
    \sigma_{\mathcal{D}}^2 = \frac{1}{2\pi}\int \mathrm{d}\ell\,\ell \,C^{\mathcal{DD}}(\ell).
\end{equation}
The higher the mean redshift of the sample, the more important does the first term in Equation~(\ref{eq:noise_level_dm}) become. This term can be interpreted as the field variance, introducing an irreducible variance due to the stochastic nature of the field. We note that the cross-power spectrum does not have any noise component since the shot noise in galaxies and the stochastic DM variance are independent and uncorrelated sources of noise, and hence their cross-correlation vanishes.

Following the general formula in Equation~\ref{eqn:generalized_covariance}, the covariance matrices for the measurement of $C^{\mathcal{DD}}(\ell)$, $C^{\mathrm{gg}}(\ell)$ and $C^{\mathrm{g}\mathcal{D}}(\ell)$ can be written as
\begin{equation}
    \mathrm{Cov}_{\mathcal{DD}}[\ell_i, \ell_j] = \frac{2 \delta^\mathrm{K}_{ij} \left(C^{\mathcal{DD}}(\ell_i) + N_\mathcal{D}\right)^2}{(2\ell_i+1)\Delta\ell f_\mathrm{sky}},
\end{equation}
\begin{equation}
    \mathrm{Cov}_{\mathrm{gg}}[\ell_i, \ell_j] = \frac{2 \delta^\mathrm{K}_{ij} \left(C^{\mathrm{gg}}(\ell_i) + N_\mathrm{g}\right)^2}{(2\ell_i+1)\Delta\ell f_\mathrm{sky}},
\end{equation}
\begin{equation}
\begin{aligned}
    \mathrm{Cov}_{\mathrm{g}\mathcal{D}}[\ell_i, \ell_j] = & \frac{\delta^\mathrm{K}_{ij}}{(2\ell_i+1)\Delta\ell f_\mathrm{sky}} [(C^{\mathrm{g}\mathcal{D}}(\ell_i))^2 \\
    & + \left(C^{\mathcal{DD}}_{\mathrm{obs}}(\ell_i) + N_\mathcal{D} \right) \left(C^{\mathrm{gg}}_{\mathrm{obs}}(\ell_i) + N_\mathrm{g}\right)].
\end{aligned}
\end{equation}
For the joint analysis of the galaxy-DM cross-power spectrum and the DM auto-power spectrum (g-DM + DM-DM), the data vector is $\left(C^{\mathrm{g}\mathcal{D}}(\ell), C^{\mathcal{DD}}(\ell)\right)^T$,
and the corresponding covariance matrix is
\begin{equation}
    \mathrm{Cov}_{\mathrm{g}\mathcal{D} + \mathcal{DD}} = 
    \begin{pmatrix}
    \mathrm{Cov}_{\mathrm{g}\mathcal{D}}[\ell_i, \ell_j] & \mathrm{Cov}_{\mathrm{g}\mathcal{D}-\mathcal{DD}}[\ell_i, \ell_j] \\
    \mathrm{Cov}_{\mathrm{g}\mathcal{D}-\mathcal{DD}}[\ell_i, \ell_j] & \mathrm{Cov}_{\mathcal{DD}}[\ell_i, \ell_j]
    \end{pmatrix},
\end{equation}
where the g-DM and DM-DM cross-variance (off-diagonal) term is
\begin{equation}
    \mathrm{Cov}_{\mathrm{g}\mathcal{D}-\mathcal{DD}}[\ell_i, \ell_j] = \frac{2 \delta^\mathrm{K}_{ij} C^{\mathrm{g}\mathcal{D}}(\ell_i) \left(C^{\mathcal{DD}}_{\mathrm{obs}}(\ell_i) + N_\mathcal{D}\right)}{(2\ell_i+1)\Delta\ell f_\mathrm{sky}}.
\end{equation}
\begin{table*}
    \centering
    \begin{tabular}{llllll}
        \toprule
        Survey & Area $f_\mathrm{sky}$ & Number of Objects & Intrinsic Noise & Tomographic Bins & Redshift Distribution \\
        \hline

        FRB & 0.35 & $10^4$ & 90~pc\,cm$^{-3}$ & 1 & $\mathrm{d}n_\mathrm{f,2D}/\mathrm{d}z \propto z^2 \exp{(-\alpha z)}$, $\alpha = 3.5$ \\

        Galaxy & 0.36 & $8 \times 10^6$ & 320~deg$^{-2}$ & 2 & $\mathrm{d}n_\mathrm{g,2D}/\mathrm{d}z \propto \bar{n} dV/\mathrm{d}z$, $\bar{n} = 5 \times 10^{-4}~h^3$Mpc$^{-3}$ \\

        \hline
    \end{tabular}
    \caption{The default settings of FRB and luminous red galaxy (LRG) surveys assumed in this work. The FRB survey settings are characteristic of upcoming surveys, such as CHIME Outriggers~\citep{2025arXiv250405192F}, CHORD~\citep{2019clrp.2020...28V}, DSA-2000~\citep{2019BAAS...51g.255H} and SKA~\citep{2004NewAR..48..979C}, and the galaxy survey settings are representative of the LRG sample from DESI survey~\citep{2023AJ....165...58Z}.}
    \label{table:survey_settings}
\end{table*}
Similarly, for the joint analysis involving the galaxy auto-power spectrum, the galaxy-DM cross-power spectrum, and the DM auto-power spectrum (g-g + g-DM + DM-DM), the data vector is $\left(C^{\mathrm{gg}}(\ell), C^{\mathrm{g}\mathcal{D}}(\ell), C^{\mathcal{DD}}(\ell)\right)^T$,
and the corresponding covariance matrix is
\begin{equation}
\begin{aligned}
& \mathrm{Cov}_{\mathrm{gg}+\mathrm{g}\mathcal{D} + \mathcal{DD}} = \\
& \begin{pmatrix}
    \mathrm{Cov}_{\mathrm{gg}}[\ell_i, \ell_j] & \!\!\mathrm{Cov}_{\mathrm{gg}-\mathrm{g}\mathcal{D}}[\ell_i, \ell_j] &\!\! \mathrm{Cov}_{\mathrm{gg}-\mathcal{DD}}[\ell_i, \ell_j] \\
    \!\mathrm{Cov}_{\mathrm{gg}-\mathrm{g}\mathcal{D}}[\ell_i, \ell_j] & \mathrm{Cov}_{\mathrm{g}\mathcal{D}}[\ell_i, \ell_j] & \!\!\mathrm{Cov}_{\mathrm{g}\mathcal{D}-\mathcal{DD}}[\ell_i, \ell_j] \\
\!    \mathrm{Cov}_{\mathrm{gg}-\mathcal{DD}}[\ell_i, \ell_j] &\!\! \mathrm{Cov}_{\mathrm{g}\mathcal{D}-\mathcal{DD}}[\ell_i, \ell_j] & \!\!\mathrm{Cov}_{\mathcal{DD}}[\ell_i, \ell_j]
\end{pmatrix}\!,
\end{aligned}
\end{equation}
where the g-g and g-DM cross-variance is
\begin{equation}
    \mathrm{Cov}_{\mathrm{gg}-\mathrm{g}\mathcal{D}}[\ell_i, \ell_j] = \frac{2 \delta^\mathrm{K}_{ij} \left(C^{\mathrm{gg}}(\ell_i) +N_\mathrm{g}\right) C^{\mathrm{g}\mathcal{D}}(\ell_i)}{(2\ell_i+1)\Delta\ell f_\mathrm{sky}},
\end{equation}
and the g-g and DM-DM cross-variance is
\begin{equation}
    \mathrm{Cov}_{\mathrm{gg}-\mathcal{DD}}[\ell_i, \ell_j] = \frac{2 \delta^\mathrm{K}_{ij} \left(C^{\mathrm{g}\mathcal{D}}(\ell_i)\right)^2}{(2\ell_i+1)\Delta\ell f_\mathrm{sky}}.
\end{equation}

\subsection{Signal-to-Noise Ratio and Fisher Forecasts} \label{subsec:signal_to_noise_ratio_and_fisher_matrix_formalism}

We forecast the expected signal-to-noise ratio (SNR) and constraints on various parameters using the Fisher information matrix. The SNR of the clustering signal from cross-power spectrum of two Gaussian fields $A$ and $B$ can be computed as
\begin{equation}
    \mathrm{SNR}^2 = \sum_{i,j} C^{AB}(\ell_i) \mathrm{Cov}^{-1}_{AB}[\ell_i, \ell_j] C^{AB}(\ell_j).
\end{equation}
Next, under the assumption of a Gaussian likelihood, the Fisher information matrix, constructed as
\begin{equation}
    F_{ij} = \frac{\partial C^{AB} (\ell)}{\partial p_i} \mathrm{Cov}^{-1}_{AB}[\ell] \left(\frac{\partial C^{AB} (\ell)}{\partial p_j}\right)^T,
\end{equation}
encodes the ability to constrain several parameters, where a large Fisher matrix means that small changes in parameters lead to large changes in the log-likelihood, thus implying good constraints. The covariance on the estimate of the parameters can then be computed as
\begin{equation}
    \langle \Delta p_i \Delta p_j \rangle = (F^{-1})_{ij}.
\end{equation}

\section{Simulation Specifications} \label{sec:simulation_specifications}

In this section, we discuss the FRB and galaxy survey settings, the power spectrum simulations and how they impact the three statistics discussed in Section~\ref{subsec:one_point_two_point_statistics}. In Section~\ref{subsec:survey_settings}, we discuss the survey specifications, the expected shot noise contributions to the auto-correlation statistics and the corresponding limits on multipoles. Then, we present our halo model-based simulations of the auto- and cross-correlation power spectra, and discuss the impact of baryonic feedback, cosmology and HOD in Section~\ref{subsec:power_spectra_simulations}.

\subsection{Survey Settings} \label{subsec:survey_settings}

We conduct SNR and Fisher forecasts for FRB and galaxy surveys described in Table~\ref{table:survey_settings}. For our baseline FRB survey, we assume discovery of $10^4$ (and $10^5$) FRB sources over 35\% of the sky. We assume an FRB redshift distribution $dn_\mathrm{f,2D}/dz \propto z^2 \exp{(-\alpha z)}$, where the exponential function represents the decreasing FRB detection efficiency (controlled by $\alpha$) at higher redshifts due to FRB luminosity function, instrument selection effects and/or intrinsic evolution of the population. We use $\alpha=3.5$, which is representative of the next-generation FRB experiments, including CHIME Outriggers~\citep{2025arXiv250405192F}, CHORD~\citep{2019clrp.2020...28V}, DSA-2000~\citep{2019BAAS...51g.255H} and SKA~\citep{2004NewAR..48..979C}. 

Large-scale structure galaxy surveys aim to build samples of galaxy populations while understanding their selection functions to allow comparisons between various samples. Of specific interest are LRGs, emission line galaxies (ELGs) and quasi-stellar objects (QSOs). In this work, we use LRGs as an example, which are bright galaxies with a prominent Balmer break, and are highly biased tracers of the large-scale structure~\citep{2023AJ....165...58Z}. Therefore, for galaxies, we assume survey specifications for the recent DESI-LRG sample, which includes 8 million objects over 36\% of the sky, with redshift distribution $dn_\mathrm{g,2D}/dz \propto \bar{n} dV/dz$ and comoving number density $\bar{n} = 5 \times 10^{-4}~h^3$Mpc$^{-3}$~\citep{2023AJ....165...58Z}.

\begin{figure}
\centering
\includegraphics[width=\columnwidth]{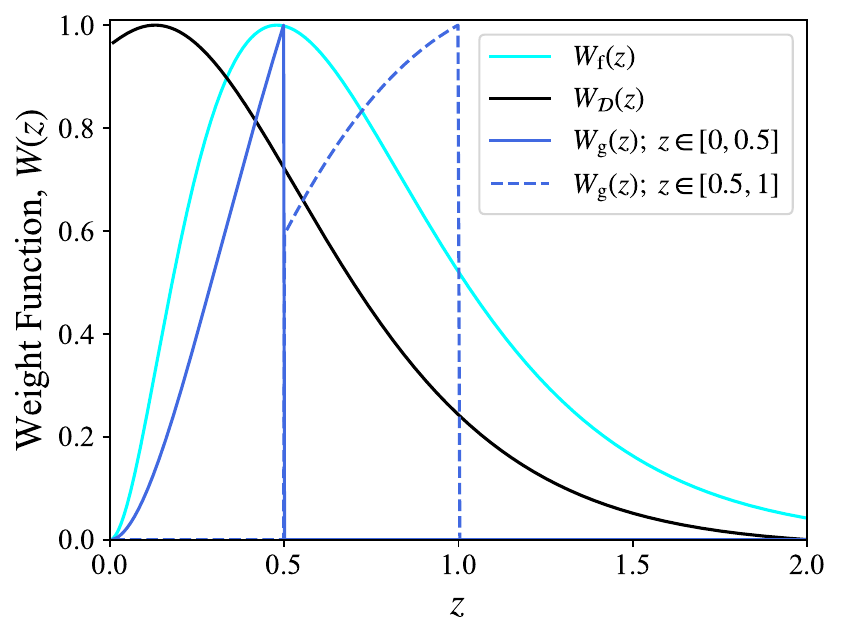}
\caption{Weighting function for FRB sources (cyan curve), FRB dispersion measures (black curve) and luminous red galaxies (LRGs, blue curves) in two tomographic bins: $z\in[0.01, 0.5]$ (solid curve) and $z\in(0.5, 1]$ (dashed curve). All weighting functions are normalized to the peak value. Although FRB sources are distributed over a broad redshift range, their dispersion measures are most sensitive to density fluctuations in the intervening foreground structure.}
\label{fig:weight_functions}
\end{figure}

For FRBs and galaxies, we assume one and two tomographic bins, respectively. Specifically, for $W_{\mathrm{g}}(z)$, the two redshift bins are $z \in [0.01, 0.5]$ and $z \in (0.5, 1]$. We visualize the weighting functions for: (i) galaxies $W_{\mathrm{g}}(z)$, (ii) FRB sources $W_{\mathrm{f}}(z)$, and (iii) DM perturbation $W_{\mathcal{D}}(z)$, in Figure~\ref{fig:weight_functions}. For illustration purposes, the weighting functions in this figure are normalized. Given the broad FRB redshift distribution discussed above, the $W_{\mathrm{f}}(z)$ is wide and peaks at intermediate redshifts $z \sim 0.6$. On the contrary, the $W_{\mathcal{D}}(z)$ peaks at lower redshifts $z \sim 0.2$. This is because, in analogy to weak lensing, the dispersion measure accrues signal from the foreground of FRB sources. This motivates the need to use multiple redshift bins for galaxies that we are cross-correlating with to avoid washing out the signal. 

Given the galaxy sample size and comoving number density, the on-sky galaxy number density is 320~deg$^{-2}$, where the split between the two tomographic bins is 75~deg$^{-2}$ and 245~deg$^{-2}$ for $z\in[0.01, 0.5]$ and $z\in(0.5, 1]$, respectively. The corresponding galaxy shot noise levels in the two tomographic bins are $4\times 10^{-6}$ and $10^{-6}$, respectively. At such low shot noise levels, scales with multipoles $\ell \leq 5000$ are accessible and not dominated by shot noise. However, due to the limitations of HoD modeling at small scales, we limit our analysis to $\ell \leq 500$. Similarly, for the aforementioned FRB survey specifications, the on-sky FRB source number density is 0.4~deg$^{-2}$ (4~deg$^{-2}$) and the mean FRB source redshift is 0.6. Assuming variance in host dispersion measure $\sigma_{\mathrm{host}} = 90$~pc\,cm$^{-3}$~\citep{2024arXiv240916952C}, the shot noise level for FRB DMs is 8~pc$^2$cm$^{-6}$ for $10^4$ FRBs and 0.7~pc$^2$cm$^{-6}$ for $10^5$ FRBs, which limits accessible multipoles to $\ell \lesssim 20$ and $\ell \lesssim 100$, respectively. For these multipole limits, the accessible angular scales are $\gtrsim$degree-scale. Therefore, in our analysis, we limit multipoles in the range $\ell \in [1, 500]$ with binning $\Delta \ell = 1$ for simplicity.

\begin{figure*}[ht!]
\centering
\includegraphics[width=\textwidth]{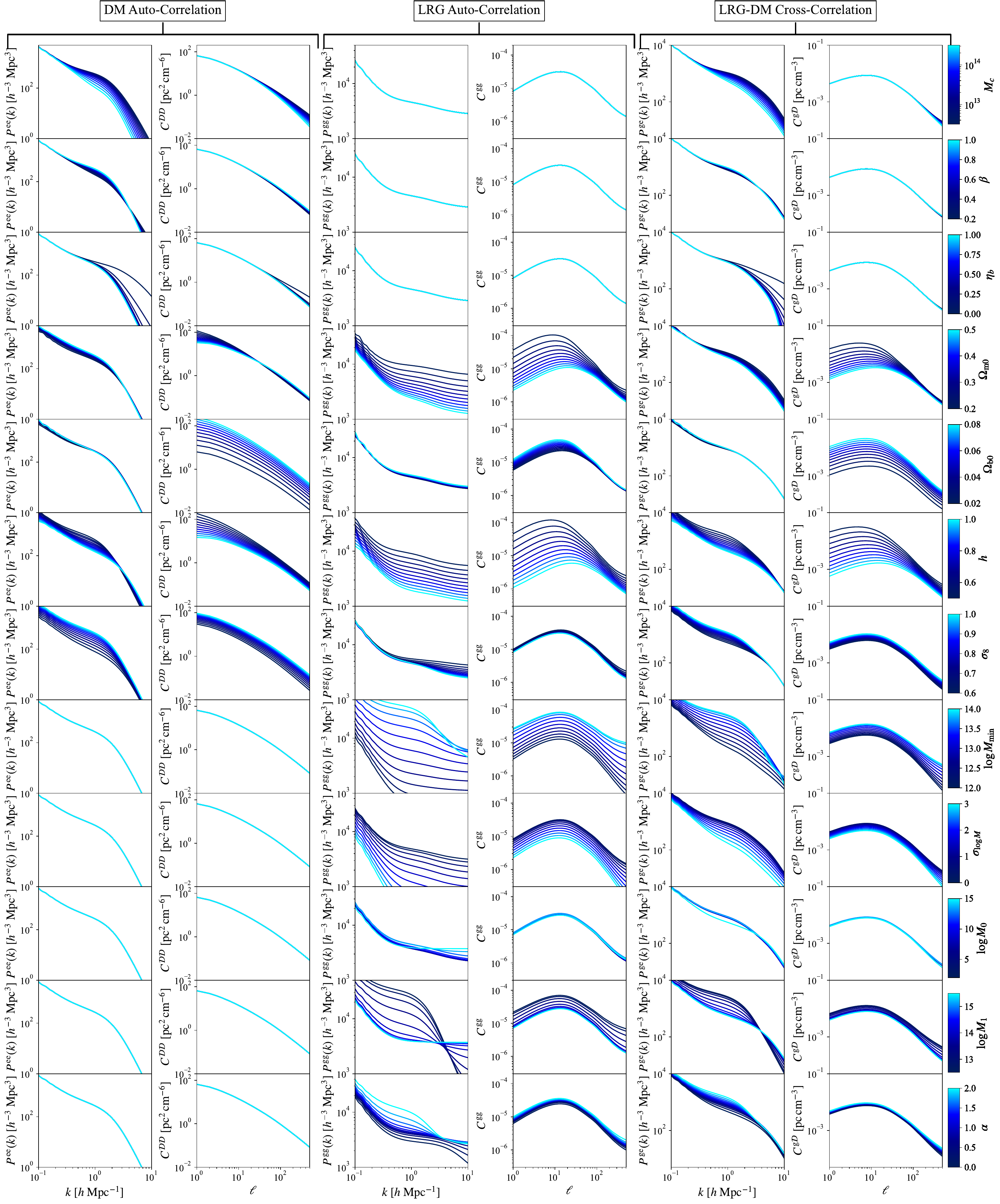}
\caption{The impact of baryonic feedback strength ($M_0$, $\beta$,  $\eta_b$), cosmological parameters ($\Omega_\mathrm{m0}$, $\Omega_\mathrm{b0}$, $h$, $\sigma_8$) and LRG halo occupation distribution ($\log M_\mathrm{min}$, $\sigma$, $\log M_0$, $\log M_1$, $\alpha$) on the physical quantities and observables for FRB dispersion measures auto-correlation ($P^\mathrm{ee}, C^\mathcal{DD}$), LRG auto-correlation ($P^\mathrm{gg}, C^\mathrm{gg}$) and LRG-dispersion measure cross-correlation ($P^\mathrm{ge}, C^{\mathrm{g}\mathcal{D}}$). As expected, the $C^\mathcal{DD}$ ($C^\mathrm{gg}$) do not vary with the LRG halo occupation distribution (baryonic feedback).}
\label{fig:feedback_variations}
\end{figure*}

\subsection{Power Spectrum Simulations} \label{subsec:power_spectra_simulations}

\begin{figure*}[ht!]
\centering
\includegraphics[width=\textwidth]{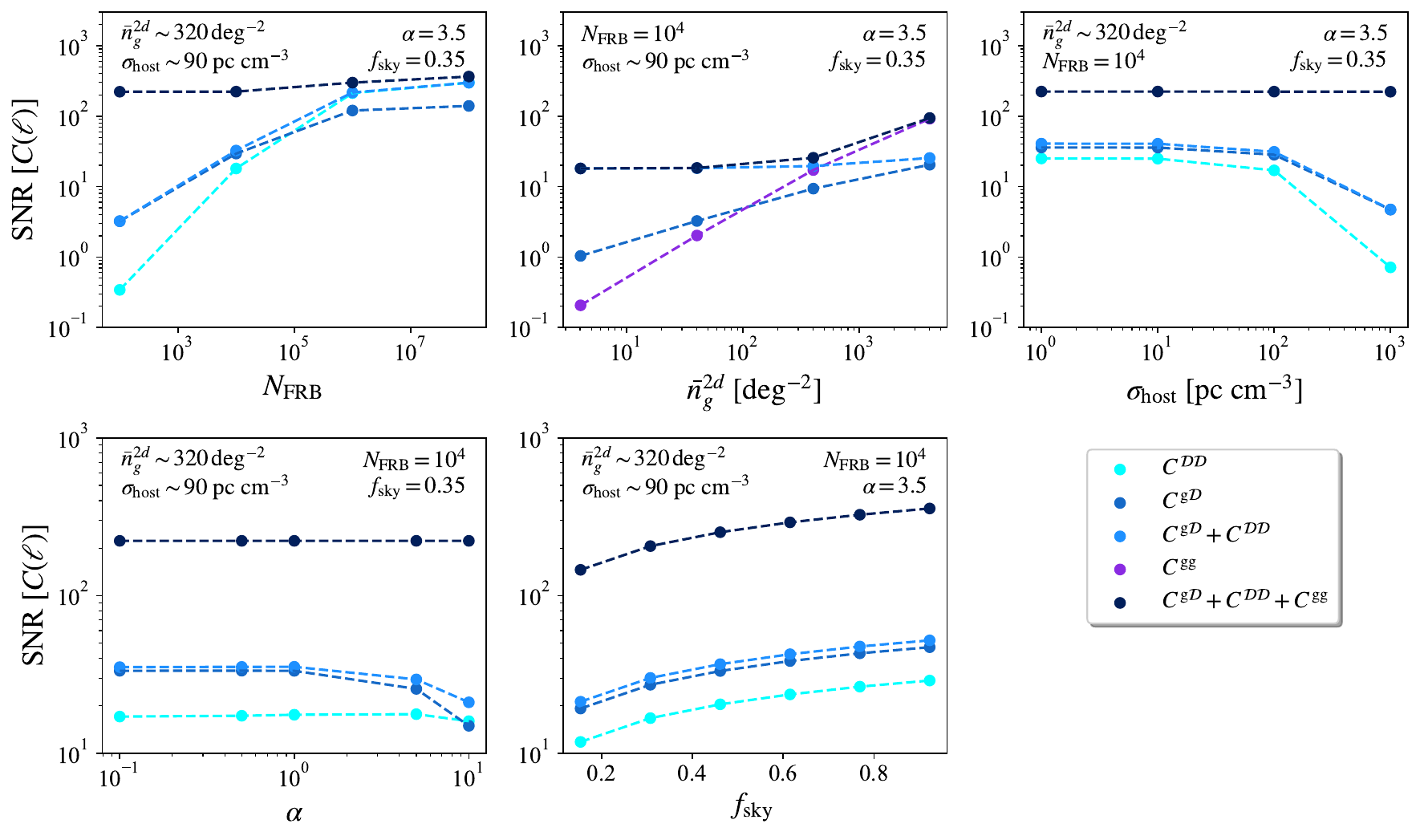}
\caption{Signal-to-noise ratio (SNR) forecasts of auto/cross-correlation observables as a function of FRB sample size ($N_\mathrm{FRB}$), mean galaxy number density ($\bar{n}_\mathrm{g}^\mathrm{2d}$), uncertainty in host dispersion measure ($\sigma_\mathrm{host}$), FRB redshift distribution parameter ($\alpha$) and fraction of sky covered by the FRB survey ($f_\mathrm{sky}$). The fiducial value of fixed parameters is indicated on the plot.}
\label{fig:snr}
\end{figure*}

To simulate variations of power spectra with baryonic feedback strength, cosmology and galaxy HOD parameters, we use the halo model prescription. Specifically, for gas profiles, we use the \citet{2020A&A...641A.130M} halo model, as implemented in \sw{BaryonForge}\footnote{https://github.com/DhayaaAnbajagane/BaryonForge}~\citep{2024OJAp....7E.108A}. The three feedback parameters, controlling the bound gas fraction in halos (see Equation~25 and 35 of \citet{2020A&A...641A.130M} for reference), include (i) $M_c$: the halo mass below which halos have lost more than 50\% of their initial gas content, (ii) $\beta$: the low-mass power-law slope of the halo bound gas fraction, and (iii) $\eta_b$: controls the ejected gas radius. For halo galaxy number density profile, we use the \sw{pyccl} software package\footnote{https://github.com/LSSTDESC/CCL}, which is based on the \citet{2007ApJ...667..760Z} HOD model parameters (see Equation~4 and 5 of \citet{2024MNRAS.530..947Y} for reference), namely (i) $\log M_\mathrm{min}$, which sets the minimum halo mass that can host a central galaxy, (ii) $\sigma_{\log M}$, which controls the steepness of the transition in the number of central galaxies from 0 to 1, (iii) $\log M_0$, which sets the minimum halo mass that can host a satellite galaxy, (iv) $\log M_1$, which sets the typical halo mass that can host a satellite galaxy, and (v) $\alpha$, which is the power law index on the number of satellite galaxies. The four cosmological parameters used in our analysis, governing the shape/amplitude of the power spectra at various scales, include (i) $\Omega_\mathrm{m0}$, (ii) $\Omega_\mathrm{b0}$, (iii) $h = H_0/100$, and (iv) $\sigma_8$: the root-mean-square amplitude of the linear matter density fluctuations smoothed with a top-hat filter of comoving radius 8\,$h^{-1}$Mpc.

We illustrate the impact of baryonic feedback, cosmology and HOD parameters on power spectra in Figure~\ref{fig:feedback_variations}. As expected, feedback strongly impacts the $P_\mathrm{ee}(k)$ on small scales ($k \gtrsim 0.5~h$\,Mpc$^{-1}$), and consequently, the $C^\mathcal{DD}(\ell)$ at large multipoles ($\ell \gtrsim 10$). Although feedback has no impact on $P_\mathrm{gg}(k)$ and $C^\mathrm{gg}(\ell)$, both quantities correlate strongly with cosmology and HOD parameters. We note that the lack of impact of feedback on $P_\mathrm{gg}(k)$ is a modeling choice when using the HoDs, where the parameters are not linked to baryon model parameters. On the contrary, $P_\mathrm{ge}(k)$ and $C^{\mathrm{g}\mathcal{D}}(\ell)$ are sensitive to all feedback, cosmology and galaxy HOD parameters.

\section{Results and Discussion} \label{sec:results_discussion}

\setlength{\tabcolsep}{4pt}
\begin{table*}
    \centering
    \begin{tabular}{lrrrrrrrrrrrrr}
        \toprule
        Observable & $\log M_c$ & $\beta$ & $\eta_b$ & $\Omega_\mathrm{m0}$ & $\Omega_\mathrm{b0}$ & $h$ & $\sigma_8$ & $\log M_\mathrm{min}$ & $\sigma$ & $\log M_0$ & $\log M_1$ & $\alpha$ & SNR \\
        
        \hline

        $C^\mathcal{DD}$ & 8.20 & 99.76 & 99.86 & 83.05 & 62.56 & 65.77 & 63.70 & 100.00 & 100.00 & 100.00 & 100.00 & 100.00 & 18.07 \\

        & 6.11 & 99.21 & 99.28 & 75.05 & 49.25 & 49.33 & 49.43 & 100.00 & 100.00 & 100.00 & 100.00 & 100.00 & 79.87 \\

        $C^{\mathrm{g}\mathcal{D}}$ & 18.24 & 99.96 & 99.98 & 82.41 & 67.07 & 54.27 & 67.17 & 4.98 & 99.93 & 99.73 & 11.20 & 99.81 & 29.40 \\

        & 8.49 & 99.82 & 99.95 & 72.37 & 52.09 & 42.61 & 54.05 & 2.59 & 99.78 & 98.86 & 8.44 & 99.73 & 71.60 \\

        $C^\mathrm{gg}$ & 100.00 & 100.00 & 100.00 & 24.11 & 21.34 & 15.49 & 21.27 & 0.37 & 96.67 & 81.97 & 4.02 & 99.14 & 221.92 \\

        ($C^{\mathrm{g}\mathcal{D}}$, $C^\mathcal{DD}$) & 5.50 & 98.70 & 99.13 & 72.81 & 43.89 & 47.10 & 44.70 & 2.72 & 99.23 & 98.42 & 10.09 & 99.67 & 32.48 \\

        & 3.48 & 94.58 & 96.61 & 40.97 & 30.15 & 27.14 & 30.57 & 1.50 & 97.67 & 94.92 & 7.26 & 99.39 & 95.81 \\

        ($C^{\mathrm{g}\mathcal{D}}$, $C^\mathcal{DD}$, $C^\mathrm{gg}$) & 3.42 & 94.20 & 97.57 & 17.54 & 10.72 & 12.52 & 10.65 & 0.21 & 93.19 & 70.73 & 3.74 & 87.61 & 222.58 \\

        & 2.46 & 83.92 & 90.93 & 12.28 & 7.52 & 8.83 & 7.64 & 0.16 & 90.18 & 65.63 & 3.48 & 77.71 & 233.92 \\

        \hline
    \end{tabular}
    \caption{Marginalized Fisher parameter constraint forecasts for feedback, cosmological, and halo occupation distribution parameters, assuming a Gaussian prior on parameter $p$ with standard deviation equal to its fiducial value: $\sigma(p) = p$. The percentage constraints are computed as $\sigma(p)/p \times 100$, and therefore, a constraint of 100\% implies that the prior is recovered. The first (second) line for each observable corresponds to constraints with $10^4$ ($10^5$) FRBs. As expected, FRB (DM-DM) and galaxy (g-g) auto-correlations, alone, cannot constrain halo occupation distribution and feedback, respectively. While $N_\mathrm{FRB} = 10^4$ offers reasonable constraints on all parameters through DM-DM or g-DM correlations, their constraining power can be improved by combining the two observables (g-DM + DM-DM), while accounting for their cross-variance. The best constraining power can be achieved with g-DM + DM-DM + g-g joint analysis as a 3$\times$2-point statistic.}
    \label{table:fisher_constraints}
\end{table*}

In this section, we present our SNR (see Section~\ref{subsec:snr_forecasts}) and Fisher parameter constraint (see Section~\ref{subsec:fisher_constraints}) forecasts for all statistics described in Section~\ref{sec:formalism}. Specifically, we emphasize the importance of combining the three angular power spectra as a 3$\times$2-point statistic to yield the strongest constraints on astrophysical feedback, as well as cosmology.

\subsection{Signal-to-Noise Ratio Forecasts} \label{subsec:snr_forecasts}

Following the statistical formalism in Section~\ref{sec:formalism}, we first report our SNR forecasts of all correlation statistics (see Table~\ref{table:fisher_constraints} for a summary) for survey settings described in Table~\ref{table:survey_settings}. Due to the low shot noise level for the galaxy survey, the expected SNR of $C^\mathrm{gg}(\ell)$ is high (SNR $\sim 220$). On the contrary, since most of the scales are dominated by shot noise for FRBs (8~pc$^2$cm$^{-6}$ for $10^4$ FRBs), the SNR of $C^\mathcal{DD}(\ell)$ and $C^{\mathrm{g}\mathcal{D}}(\ell)$ are relatively lower (SNR $\sim 18$ and $\sim 30$, respectively). The overall SNR of FRB observables can be improved through a joint analysis of $(C^{\mathrm{g}\mathcal{D}}(\ell), C^\mathcal{DD}(\ell))^T$, thus yielding an SNR $\sim 32$. The highest SNR of $\sim 220$ can be achieved through a joint analysis of FRB and galaxy observables $(C^{\mathrm{gg}}(\ell), C^{\mathrm{g}\mathcal{D}}(\ell), C^\mathcal{DD}(\ell))^T$.

We examine the impact of several survey parameters on the SNR of clustering signal in Figure~\ref{fig:snr}. As noted in Equation~\ref{eq:noise_level_dm}, the shot noise depends on the mean FRB source angular number density $\bar{n}_{\mathrm{f,2D}}$, the mean redshift of the observed FRB sample $\bar{z}_\mathrm{f}$ and the variance in host DM contribution $\sigma^2_\mathrm{host}$. The $\bar{n}_{\mathrm{f,2D}}$ depends on the total number of FRBs discovered $N_\mathrm{FRB}$ and the fraction of sky covered by the survey $f_\mathrm{sky}$. Furthermore, covariance also explicitly depends on $f_\mathrm{sky}$ (see Section~\ref{subsec:covariance_matrices} and Equation~\ref{eqn:generalized_covariance}). Therefore, as expected, we observe a rapid increase in SNR of the clustering signal as $N_\mathrm{FRB}$ and $f_\mathrm{sky}$ increases. The $\bar{z}_\mathrm{f}$ depends on the redshift sensitivity of the survey, controlled by $\alpha$. Therefore, the SNR of the clustering signal decreases as the redshift sensitivity of the survey decreases ($\alpha$ increases). However, at $\alpha \lesssim 1$, the signal detection significance saturates because the galaxies that we are cross-correlating with are located at redshifts $z_\mathrm{g} \leq 1$. The detection significance also improves with decreasing $\sigma_\mathrm{host}$. However, the differential improvement in detection significance with decreasing $\sigma_\mathrm{host}$ begins to saturate at $\sigma_\mathrm{host} \lesssim 100$~pc\,cm$^{-3}$ because the field variance (see Equation~\ref{eq:noise_level_dm}), which is $\sim 75$~pc\,cm$^{-3}$, starts to dominate the shot noise.

\begin{figure*}
\centering
\includegraphics[width=\textwidth]{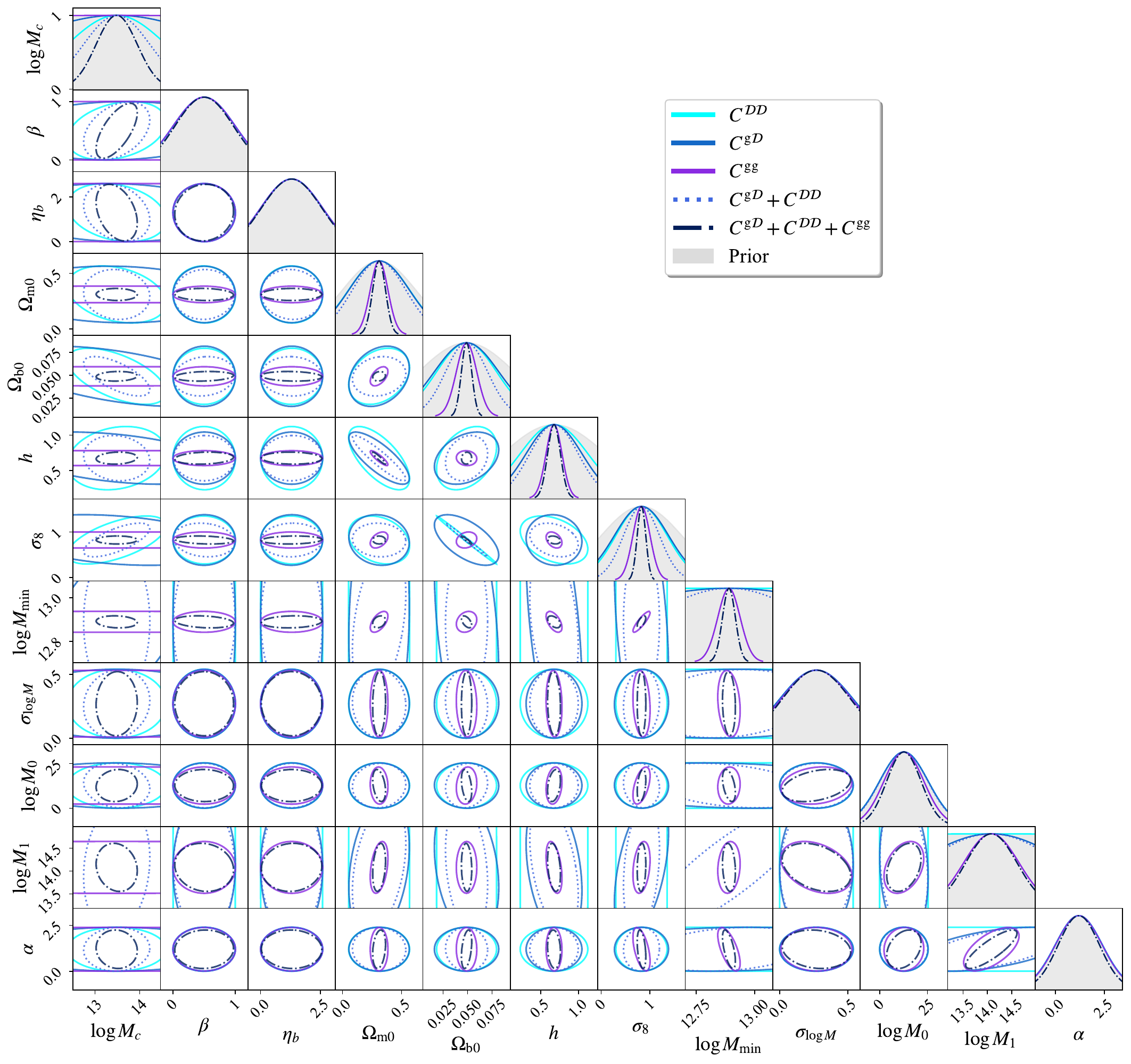}
\caption{Fisher matrix forecasts for FRB and galaxy surveys. The ellipses for 68\% confidence intervals from several observables are shown: $C^\mathcal{DD}$ (cyan curves), $C^{\mathrm{g}\mathcal{D}}$ (blue curves), $C^{\mathrm{gg}}$ (violet curves), $C^{\mathrm{g}\mathcal{D}} + C^\mathcal{DD}$ (dotted curves) and $C^{\mathrm{g}\mathcal{D}} + C^\mathcal{DD} + C^{\mathrm{gg}}$ (dash-dot curves). While $C^{\mathrm{gg}}$ outperforms $C^\mathcal{DD}$ and $C^{\mathrm{g}\mathcal{D}}$ in constraining cosmology, it lacks constraining power on the strength of baryonic feedback. The $C^{\mathrm{g}\mathcal{D}} + C^\mathcal{DD} + C^{\mathrm{gg}}$ joint analysis yields the most stringent constraints overall.}
\label{fig:constraints}
\end{figure*}

\begin{figure*}
\centering
\includegraphics[width=0.7\textwidth]{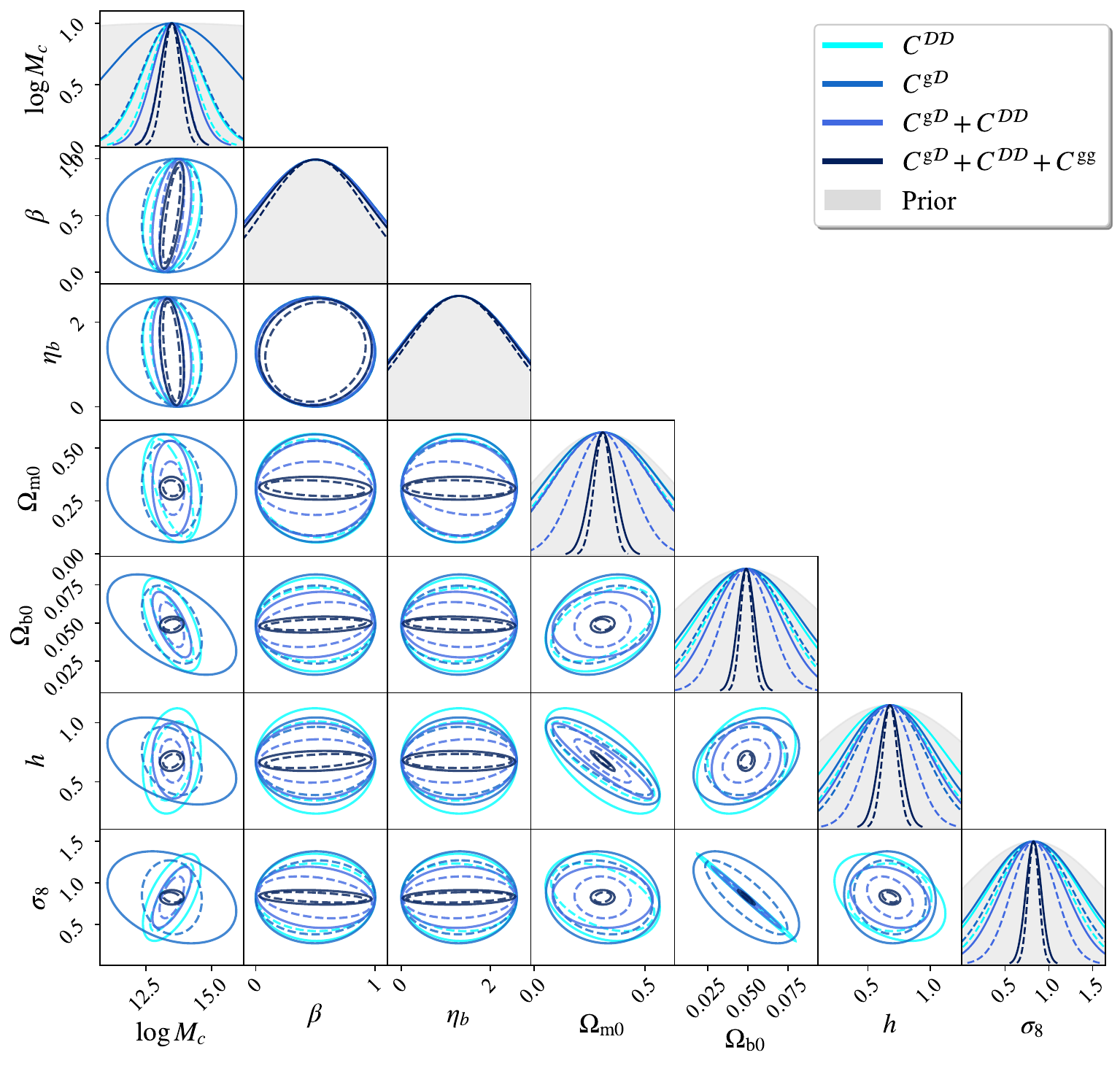}
\caption{Fisher matrix forecasts (same as Figure~\ref{fig:constraints}), marginalized over galaxy halo occupancy distribution, for galaxy surveys, combined with $10^4$ (solid lines) and $10^5$ (dashed lines) FRBs. The improvement in percentage constraints from the 3$\times$2-point statistic $C^{\mathrm{g}\mathcal{D}} + C^\mathcal{DD} + C^{\mathrm{gg}}$ by increasing the FRB sample size by a factor of 10 is $\sim 25$\% on $\log M_c$, $\sim 12$\% on $\beta$, $\sim 8$\% on $\eta_b$ and $\sim 30$\% on all cosmological parameters.}
\label{fig:constraints_more_Nfrb}
\end{figure*}

\begin{figure*}[ht!]
\centering
\includegraphics[width=\textwidth]{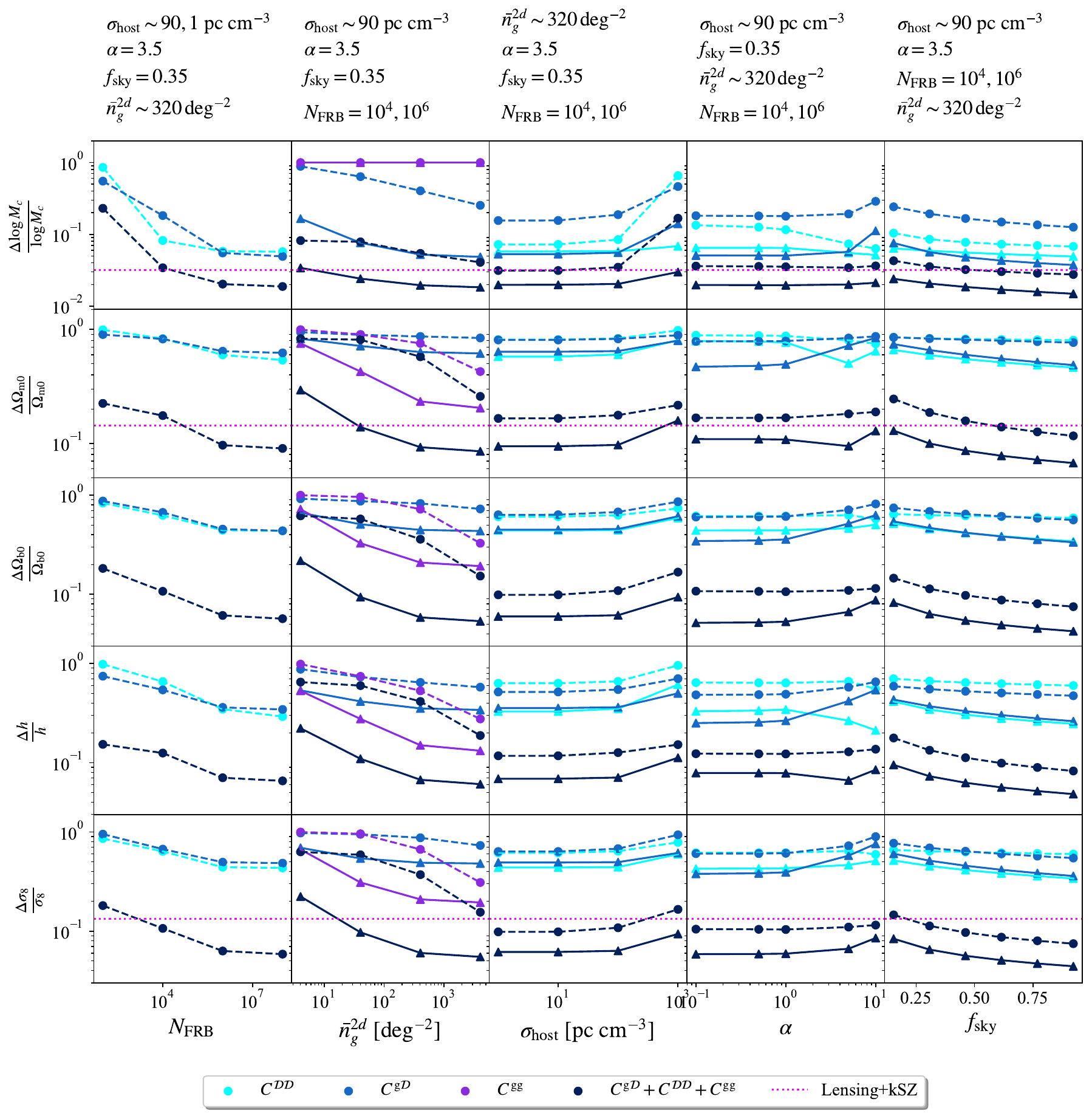}
\caption{Fisher matrix forecasts of cosmological and feedback parameters for auto/cross-correlation observables as a function of survey parameters. The fixed parameters for each column are listed on the top. We show variations with $N_\mathrm{FRB} = 10^4$ (dashed lines) and $N_\mathrm{FRB} = 10^6$ (solid lines) For comparison, we show  constraints from weak lensing and kSZ effect~\citep{2024MNRAS.534..655B} as pink dotted lines. The expected constraints from FRBs are competitive with other probes of baryonic feedback.}
\label{fig:parameter_constraints_vary_survey}
\end{figure*}

\subsection{Fisher Constraints} \label{subsec:fisher_constraints}

We report the Fisher parameter constraint forecasts for feedback, cosmology and HOD parameters for cross-correlation observables, following the formalism in Section~\ref{sec:formalism} for survey settings summarized in Table~\ref{table:survey_settings}. These constraints are computed by assuming a Gaussian prior on parameter $p$ with standard deviation $\sigma(p) = p$ and therefore, a 100\% constraint implies that the prior is recovered (no constraining power). The one-dimensional marginal percentage constraints on all parameters are summarized in Table~\ref{table:fisher_constraints} and the corresponding corner plot is shown in Figure~\ref{fig:constraints}. The prior is shown as shaded area in the one-dimensional histograms for reference. 

For the DM auto-correlation statistic $C^{\mathcal{DD}}$, we observe that it reasonably constrains the feedback parameter $\log M_c$ and all cosmological parameters, and recovers the underlying prior for $\beta$, $\eta_b$ and HOD parameters (as expected). The poor constraining power on feedback parameters $\beta$ and $\eta_b$ is because these parameters impact relatively smaller scales ($\ell \gtrsim 100$), which are shot noise dominated for the presumed survey settings. Additionally, we observe that feedback is strongly degenerate with cosmology, where $\log M_c$ negatively correlates with $\Omega_\mathrm{m0}$ and $\Omega_\mathrm{b0}$, and positively correlates with $h$ and $\sigma_8$. Since $\log M_c$ sets the mass scale below which halos are more than 50\% evacuated of gas, increasing $\log M_c$ means that feedback is more efficient. Therefore, $\log M_c$ is negatively correlated with $\Omega_{\mathrm{m}0}$ and $\Omega_{\mathrm{b}0}$, since a higher matter or baryon density naturally enhances the DM signal, thereby requiring less efficient feedback to match observations. Conversely, $\log M_c$ shows a positive correlation with $h$ and $\sigma_8$, as a larger Hubble constant implies a smaller comoving distance to a given redshift, and to compensate, the model requires a larger contribution to DM, which can be achieved by increasing the efficiency of feedback and a higher $\sigma_8$ boosts small-scale structure, and to suppress the resulting structure, the model compensates by removing gas from halos. 

The $C^{\mathrm{g}\mathcal{D}}$ exhibits slightly better constraining power on $\beta$ and $\eta_b$ feedback parameters, as well as the HOD parameters, but is still impacted by limited knowledge, given the range of accessible scales. The degeneracy between feedback and cosmological parameters is weaker for $C^{\mathrm{g}\mathcal{D}}$ than for $C^{\mathcal{DD}}$, because the latter is directly sensitive to the distribution of free electrons, which is strongly modulated by feedback processes. In contrast, $C^{\mathrm{g}\mathcal{D}}$ probes how biased galaxy tracers relate to the projected electron density field. This added layer of bias, and reduced sensitivity to low-mass halos (galaxies are more likely to exist in massive halos), makes the g-DM observable less responsive to feedback-induced changes, thus reducing the strength of degeneracy with cosmological parameters. 

The complementarity of $C^{\mathrm{g}\mathcal{D}}$ and $C^{\mathcal{DD}}$ underscores the power of joint analyses in disentangling cosmological and astrophysical feedback effects. In fact, the combined analysis of $C^{\mathrm{g}\mathcal{D}}$ and $C^{\mathcal{DD}}$ enhances the constraints on $\log M_c$ by almost a factor of 2, thus highlighting the value of leveraging multiple observables. However, we emphasize that the constraining power of such joint analyses is strongly limited by the FRB shot noise contribution. The expected constraining power of a joint $C^{\mathrm{g}\mathcal{D}}$ and $C^{\mathcal{DD}}$ analysis is $\sim 6$\% on $\log M_c$, $\sim$73\% on $\Omega_\mathrm{m0}$, $\sim 44$\% on $\Omega_\mathrm{b0}$, $\sim 47$\% on $h$ and $\sim 45$\% on $\sigma_8$. Comparing these forecasts with galaxy auto-correlation $C^{\mathrm{gg}}$, we find that the upcoming galaxy surveys significantly outperform $C^{\mathrm{g}\mathcal{D}}$ and $C^{\mathcal{DD}}$ in terms of constraining power for cosmological and HOD parameters by a factor of 2-3. However, a limitation of $C^{\mathrm{gg}}$ is that it is not currently sensitive to baryonic feedback.

Motivated by the complementary nature of $C^{\mathrm{g}\mathcal{D}}$, $C^{\mathcal{DD}}$ and $C^{\mathrm{gg}}$, we assess the constraining power of 3$\times$2-point statistic joint analysis, which should allow us to constrain feedback strength, while also improving constraints on cosmological and HOD parameters. For the presumed survey settings, we forecast that the 3$\times$2-point statistic can constrain cosmological parameters, $\Omega_\mathrm{m0}$ to $\sim 18$\%, $\Omega_\mathrm{b0}$ to $\sim 11$\%, $h$ to $\sim 12$\% and $\sigma_8$ to $\sim 11$\%. Furthermore, such an analysis can also constrain feedback parameters, such as $\log M_c$ to $\sim 3$\%. These results highlight the power of multi-probe analyses in simultaneously disentangling baryonic feedback and refining cosmological parameter constraints. 

Looking forward to the next generation FRB survey designs, we also compare the expected constraints from $10^4$ FRBs with those from $10^5$ FRBs in Figure~\ref{fig:constraints_more_Nfrb}. For simplicity, we show the constraints marginalized over the halo occupation distribution function parameters. As expected, for DM-DM auto-correlations, the constraints on $\log M_c$ improve by $\sim$25\% and on cosmological parameters improve by $\sim$10-25\% (see Table~\ref{table:fisher_constraints}). Similarly, for galaxy-DM cross-correlations, the constraints on $\log M_c$ improve by $\sim 50$\% and on cosmological parameters improve by $\sim$10-25\%. The most stringent constraints are obtained from 3$\times$2-point statistic, where the differential improvement in percentage constraints is $\sim 25$\% on $\log M_c$, $\sim 12$\% on $\beta$, $\sim 8$\% on $\eta_b$ and $\sim 30$\% on all cosmological parameters.

We also compare the expected constraints on all parameters for various survey designs in Figure~\ref{fig:parameter_constraints_vary_survey} with the constraining power of Dark Energy Survey (Y3) weak lensing and Atacama Cosmology Telescope kSZ effect measurements~\citep{2024MNRAS.534..655B}. We find that the expected constraining power of FRBs is competitive with other probes of baryonic feedback. While the forecasting formalism of \citet{2025PhRvD.111d3529P} for weak lensing auto-correlations and weak lensing-tSZ cross-correlations
is distinct from the approach followed in this work, the forecasts for $\Omega_\mathrm{m}$ and $\sigma_8$ for Vera Rubin observatory and Simons observatory are also competitive with FRBs. However, we note that fair comparisons with galaxy clustering and cosmic shear measurements from upcoming surveys like Vera Rubin observatory and Euclid are not possible at the current stage as their forecasts do not incorporate the impact of complex feedback physics~\citep{2018arXiv180901669T, 2020A&A...642A.191E}. Also, in contrast to our $\Lambda$CDM forecasts, these surveys forecast for either models beyond-$\Lambda$CDM or by fixing several cosmological parameters~\citep{2018arXiv180901669T, 2020A&A...642A.191E}, thus prohibiting fair comparisons.

\section{Lessons from Weak Lensing}\label{sec:discussion}

The forecasts presented here provide a theoretical demonstration of the constraining power of 3$\times$2-point statistic of FRBs and galaxies. In practice, realizing these gains will require careful treatment of modeling uncertainties pertaining to the variance in host DM contribution $\sigma_\mathrm{host}$ and redshift distribution of FRB sources specified by $\alpha$, in much the same way that the weak lensing community has developed increasingly sophisticated pipelines to reach percent-level precision in recent years. We highlight several key considerations for implementing this methodology for FRB surveys.

First, the estimation of weighting functions for both FRBs ($W_\mathcal{D}(z)$) and galaxies ($W_\mathrm{g}(z)$) relies on accurate redshift distributions. Underestimation of photometric redshift uncertainties and catastrophic outliers has been demonstrated to bias the inferred clustering and lensing signals~\citep{2010MNRAS.401.1399B, 2025arXiv250502928T}. Analogously, the accuracy of FRB redshift distribution estimates, which are generally based on either, (i) the FRB host galaxy association, or (ii) the FRB luminosity function, is subject to our knowledge of FRB sources and their host galaxies~\citep{2018MNRAS.481.2320L, 2021ApJ...911...95A}. The telescope beam shape and burst widths are also known impact the sensitivity to high redshift FRBs and must be modeled together with the FRB luminosity function in the second approach~\citep{2022MNRAS.509.4775J, 2022MNRAS.516.4862J} for unbiased cosmological inference.

Second, calibration errors must be considered. In cosmic shear, intrinsic alignments of galaxies, multiplicative shear calibration, and PSF modeling errors can mimic or bias the lensing signal. In the FRBs case, analogous contaminants include the DM selection function of the instrument and the variance in host DM contributions. If the instrument DM selection function prohibits the detection of high DM FRBs, it can bias our feedback constraints for massive halos~\citep{2025arXiv250603258Q}. Robust survey simulations and end-to-end injection studies, already common in optical weak lensing pipelines, will be essential for quantifying these effects for FRBs~\citep{2023AJ....165..152M}. Furthermore, an underestimation of host DM variance can both, mimic weak feedback scenarios, as well as artificially increase the signal detection significance. Robust estimates of $\sigma_\mathrm{host}$ to $\sim$100~pc\,cm$^{-3}$ precision, either through one-point statistic analysis or through studies of the host galaxy properties, would be quintessential to get the best signal detection significance (see Figure~\ref{fig:parameter_constraints_vary_survey}). 

The lessons learned from two decades of weak lensing surveys provide a clear roadmap: systematic uncertainties will need to be characterized, parameterized, and marginalized over, rather than ignored. By adopting these best practices, FRB--galaxy cross-correlations can move from theoretical promise to a robust probe.

\newpage
\section{Conclusions} \label{sec:conclusion}

In this work, we investigated the potential of FRB--galaxy analyses in constraining baryonic feedback and cosmological parameters, thus highlighting the complementary role of FRBs as a probe of gas distribution in the Universe, alongside other probes. We now provide a summary of the principal outcomes of our work.

\begin{itemize}[leftmargin=*]

\item We defined the DM perturbation field $D(\hat{x}, z)$ as a key observable derived from the FRB dispersion measure (see Equation~\ref{eqn:meanDM}). This perturbation traces fluctuations in the free electron density field $\delta_\mathrm{e}(\hat{x}, z)$, and its angular power spectrum $C^{\mathcal{DD}}(\ell)$ encodes the electron auto-power spectrum $P_\mathrm{ee}(k, z)$ (see Equation~\ref{eqn:DMDM}). The redshift sensitivity of this observable is governed by a weighting function $W_\mathcal{D}(z)$, which peaks at low redshifts relative to the FRB source distribution, making the DM perturbation most sensitive to density fluctuations in the foreground of FRB sources (see Equation~\ref{eqn:mean_projected_DM_perturbations} and Figure~\ref{fig:weight_functions}). Consequently, $C^{\mathcal{DD}}(\ell)$ is directly responsive to baryonic feedback and cosmology via its dependence on $P_\mathrm{ee}(k, z)$ and $W_\mathrm{DM}(z)$ (see Equation~\ref{eqn:meanDM} and Figure~\ref{fig:feedback_variations}).

\item We defined the cross-correlation between the angular DM perturbation field $\mathcal{D}_\mathrm{2D}(\hat{x})$ and the projected galaxy over-density field $\delta_{\mathrm{g},2\mathrm{D}}(\hat{x})$ through their angular cross-power spectrum $C^{\mathrm{g}\mathcal{D}}(\ell)$ (see Equation~\ref{eqn:angular_overdensity_field}, \ref{eqn:DM_2D} and \ref{eqn:gDM}). The $C^{\mathrm{g}\mathcal{D}}(\ell)$ is sensitive to the spatial correlation between galaxies -- which are biased tracers of the underlying matter distribution -- and the free electrons that contribute to FRB DMs. This cross-correlation effectively measures the galaxy-electron cross-power spectrum $P_\mathrm{ge}(k, z)$. Since galaxies predominantly occupy massive halos, this observable probes how feedback processes influence the ejection or retention of gas in these environments. It complements the DM auto-correlation by incorporating galaxy bias and modulating sensitivity to gas in collapsed structures, thereby allowing for improved mapping of gas-galaxy co-location across cosmic scales.

\item We emphasize the complementary nature of $C^{\mathcal{DD}}$ and $C^{\mathrm{g}\mathcal{D}}$, which exhibit different degeneracy directions in feedback and cosmology parameter space. For example, in $C^{\mathcal{DD}}$, the feedback parameter $\log M_c$ is strongly anti-correlated with $\Omega_\mathrm{m0}$ and $\Omega_\mathrm{b0}$, and positively correlated with $h$ and $\sigma_8$, due to the direct sensitivity of $C^{\mathcal{DD}}$ to the integrated electron content. On the other hand, $C^{\mathrm{g}\mathcal{D}}$ involves an additional galaxy bias and is less sensitive to diffuse gas in low-mass halos, which reduces its degeneracy with cosmological parameters. Therefore, combining these two observables helps to break degeneracies that are otherwise intrinsic to either statistic alone (see Figure~\ref{fig:constraints}).

\item Although the combination of DM auto-correlation and galaxy-DM cross-correlation, i.e., $C^{\mathcal{DD}} + C^{\mathrm{g}\mathcal{D}}$, enhances sensitivity to baryonic feedback, this joint analysis is factor of two less powerful than the galaxy auto-correlation $C^{\mathrm{gg}}$ in constraining cosmology and HOD parameters (see Table~\ref{table:fisher_constraints}). This is primarily due to the high shot noise in FRB measurements, which arises from low FRB source number densities (even with planned next generation FRB surveys!) and significant variance in the DM perturbation field and host galaxy DM contribution. While $C^{\mathrm{gg}}$ achieves a high signal-to-noise ratio (SNR $\sim 220$), $C^{\mathcal{DD}}$ and $C^{\mathrm{g}\mathcal{D}}$ individually reach only $\sim 18$ and $\sim 30$, respectively, and their joint analysis improves this only modestly to $\sim 32$. Therefore, the small number of FRBs and the associated noise substantially limit the constraining power of FRB observables, especially for the feedback parameters that are only sensitive to small scales (see Figure~\ref{fig:feedback_variations}, \ref{fig:constraints} and Table~\ref{table:fisher_constraints}). 

\item While $C^{\mathrm{gg}}$ is highly sensitive to cosmological and HOD parameters, it offers no sensitivity to baryonic feedback, since it traces the clustering of galaxies and not the diffuse gas component (see Figure~\ref{fig:feedback_variations}). This limitation makes FRB observables uniquely valuable: both $C^{\mathcal{DD}}$ and $C^{\mathrm{g}\mathcal{D}}$ respond directly to changes in the spatial distribution of ionized gas, and are therefore essential for constraining models of feedback. Despite their lower signal-to-noise, FRB-based correlations provide critical leverage on feedback parameters that are inaccessible to galaxy-only analyses. This motivates the use of joint analyses that combine FRBs and galaxies, leveraging the strengths of each observable.

\item The joint analysis of galaxy auto-correlation, galaxy–DM cross-correlation, and DM auto-correlation -- i.e., the 3$\times$2-point statistic combining $C^{\mathrm{gg}}$, $C^{\mathrm{g}\mathcal{D}}$, and $C^{\mathcal{DD}}$ -- provides the strongest overall constraining power. Under fiducial survey settings with $10^4$ ($10^5$) FRBs, this combination yields marginalized $1\sigma$ constraints of approximately $3\%$ ($2\%$) on $\log M_c$ for feedback parameters and about $18\%$ ($12\%$) on $\Omega_\mathrm{m0}$, $11\%$ ($8\%$) on $\Omega_\mathrm{b0}$, $12\%$ ($9\%$) on $h$, and $11\%$ ($8\%$) on $\sigma_8$ for cosmological parameters. This analysis capitalizes on the complementarity of the three observables: $C^{\mathrm{gg}}$ dominates the cosmological and HOD constraints, while $C^{\mathcal{DD}}$ and $C^{\mathrm{g}\mathcal{D}}$ uniquely constrain the diffuse gas physics, leading to significantly tighter bounds than any subset of the observables alone.

\item We further provide insights on building an ideal FRB survey for such cross-correlation studies. For FRB surveys with $10^4$ ($10^5$) detections, the accessible multipole range is limited to $\ell \lesssim 20$ ($\ell \lesssim 100$), which corresponds to $\sim$~degree angular scales. This implies that arcmin-scale localizations should be sufficient to enable this measurement. The FRB source redshift distribution can then either be modeled in terms of their luminosity function (for example, see \citet{2025arXiv250608932W}) or measured using the FRB source -- galaxy position cross-correlations with arcsecond-level localizations. Furthermore, improving redshift sensitivity of the FRB survey (lowering down $\alpha$) has diminishing returns, especially if we are shot-noise limited and galaxies that we are cross-correlating with are at lower redshifts (see Figure~\ref{fig:snr} and \ref{fig:parameter_constraints_vary_survey}). For example, the constraints on cosmological parameters improve by $\sim 40$\% before hitting the shot-noise floor at $\alpha \lesssim 1$ (see Figure \ref{fig:parameter_constraints_vary_survey}). Lastly, increasing the number of redshift bins when conducting cross-correlations improves the SNR and parameter constraints.

\end{itemize}

In summary, we have demonstrated that FRB-based observables provide indispensable sensitivity to baryonic feedback, which is otherwise inaccessible through galaxy surveys alone. As future FRB and galaxy surveys improve in scale and precision, joint multi-tracer analyses will play a pivotal role in disentangling astrophysical feedback and cosmological parameters.

\begin{acknowledgments}
K.S. thanks Matthew McQuinn and Yunfei Wen for insightful conversations.
\end{acknowledgments}

\bibliography{manuscript}{}

\begin{thebibliography}{}
\expandafter\ifx\csname natexlab\endcsname\relax\def\natexlab#1{#1}\fi
\providecommand{\url}[1]{\href{#1}{#1}}
\providecommand{\dodoi}[1]{doi:~\href{http://doi.org/#1}{\nolinkurl{#1}}}
\providecommand{\doeprint}[1]{\href{http://ascl.net/#1}{\nolinkurl{http://ascl.net/#1}}}
\providecommand{\doarXiv}[1]{\href{https://arxiv.org/abs/#1}{\nolinkurl{https://arxiv.org/abs/#1}}}

\bibitem[{{Aggarwal} {et~al.}(2021){Aggarwal}, {Budav{\'a}ri}, {Deller}, {Eftekhari}, {James}, {Prochaska}, \& {Tendulkar}}]{2021ApJ...911...95A}
{Aggarwal}, K., {Budav{\'a}ri}, T., {Deller}, A.~T., {et~al.} 2021, \apj, 911, 95, \dodoi{10.3847/1538-4357/abe8d2}

\bibitem[{{Amon} {et~al.}(2022){Amon}, {Gruen}, {Troxel}, {MacCrann}, {Dodelson}, {Choi}, {Doux}, {Secco}, {Samuroff}, {Krause}, {Cordero}, {Myles}, {DeRose}, {Wechsler}, {Gatti}, {Navarro-Alsina}, {Bernstein}, {Jain}, {Blazek}, {Alarcon}, {Fert{\'e}}, {Lemos}, {Raveri}, {Campos}, {Prat}, {S{\'a}nchez}, {Jarvis}, {Alves}, {Andrade-Oliveira}, {Baxter}, {Bechtol}, {Becker}, {Bridle}, {Camacho}, {Carnero Rosell}, {Carrasco Kind}, {Cawthon}, {Chang}, {Chen}, {Chintalapati}, {Crocce}, {Davis}, {Diehl}, {Drlica-Wagner}, {Eckert}, {Eifler}, {Elvin-Poole}, {Everett}, {Fang}, {Fosalba}, {Friedrich}, {Gaztanaga}, {Giannini}, {Gruendl}, {Harrison}, {Hartley}, {Herner}, {Huang}, {Huff}, {Huterer}, {Kuropatkin}, {Leget}, {Liddle}, {McCullough}, {Muir}, {Pandey}, {Park}, {Porredon}, {Refregier}, {Rollins}, {Roodman}, {Rosenfeld}, {Ross}, {Rykoff}, {Sanchez}, {Sevilla-Noarbe}, {Sheldon}, {Shin}, {Troja}, {Tutusaus}, {Tutusaus}, {Varga}, {Weaverdyck}, {Yanny}, {Yin}, {Zhang}, {Zuntz}, {Aguena}, {Allam}, {Annis}, {Bacon},
  {Bertin}, {Bhargava}, {Brooks}, {Buckley-Geer}, {Burke}, {Carretero}, {Costanzi}, {da Costa}, {Pereira}, {De Vicente}, {Desai}, {Dietrich}, {Doel}, {Ferrero}, {Flaugher}, {Frieman}, {Garc{\'\i}a-Bellido}, {Gaztanaga}, {Gerdes}, {Giannantonio}, {Gschwend}, {Gutierrez}, {Hinton}, {Hollowood}, {Honscheid}, {Hoyle}, {James}, {Kron}, {Kuehn}, {Lahav}, {Lima}, {Lin}, {Maia}, {Marshall}, {Martini}, {Melchior}, {Menanteau}, {Miquel}, {Mohr}, {Morgan}, {Ogando}, {Palmese}, {Paz-Chinch{\'o}n}, {Petravick}, {Pieres}, {Romer}, {Sanchez}, {Scarpine}, {Schubnell}, {Serrano}, {Smith}, {Soares-Santos}, {Tarle}, {Thomas}, {To}, {Weller}, \& {DES Collaboration}}]{2022PhRvD.105b3514A}
{Amon}, A., {Gruen}, D., {Troxel}, M.~A., {et~al.} 2022, \prd, 105, 023514, \dodoi{10.1103/PhysRevD.105.023514}

\bibitem[{{Anbajagane} {et~al.}(2024){Anbajagane}, {Pandey}, \& {Chang}}]{2024OJAp....7E.108A}
{Anbajagane}, D., {Pandey}, S., \& {Chang}, C. 2024, The Open Journal of Astrophysics, 7, 108, \dodoi{10.33232/001c.126788}

\bibitem[{{Asgari} {et~al.}(2023){Asgari}, {Mead}, \& {Heymans}}]{2023OJAp....6E..39A}
{Asgari}, M., {Mead}, A.~J., \& {Heymans}, C. 2023, The Open Journal of Astrophysics, 6, 39, \dodoi{10.21105/astro.2303.08752}

\bibitem[{{Baptista} {et~al.}(2024){Baptista}, {Prochaska}, {Mannings}, {James}, {Shannon}, {Ryder}, {Deller}, {Scott}, {Glowacki}, \& {Tejos}}]{2024ApJ...965...57B}
{Baptista}, J., {Prochaska}, J.~X., {Mannings}, A.~G., {et~al.} 2024, \apj, 965, 57, \dodoi{10.3847/1538-4357/ad2705}

\bibitem[{{Berlind} \& {Weinberg}(2002)}]{2002ApJ...575..587B}
{Berlind}, A.~A., \& {Weinberg}, D.~H. 2002, \apj, 575, 587, \dodoi{10.1086/341469}

\bibitem[{{Bernstein} \& {Huterer}(2010)}]{2010MNRAS.401.1399B}
{Bernstein}, G., \& {Huterer}, D. 2010, \mnras, 401, 1399, \dodoi{10.1111/j.1365-2966.2009.15748.x}

\bibitem[{{Bigwood} {et~al.}(2024){Bigwood}, {Amon}, {Schneider}, {Salcido}, {McCarthy}, {Preston}, {Sanchez}, {Sijacki}, {Schaan}, {Ferraro}, {Battaglia}, {Chen}, {Dodelson}, {Roodman}, {Pieres}, {Fert{\'e}}, {Alarcon}, {Drlica-Wagner}, {Choi}, {Navarro-Alsina}, {Campos}, {Ross}, {Carnero Rosell}, {Yin}, {Yanny}, {S{\'a}nchez}, {Chang}, {Davis}, {Doux}, {Gruen}, {Rykoff}, {Huff}, {Sheldon}, {Tarsitano}, {Andrade-Oliveira}, {Bernstein}, {Giannini}, {Diehl}, {Huang}, {Harrison}, {Sevilla-Noarbe}, {Tutusaus}, {Elvin-Poole}, {McCullough}, {Zuntz}, {Blazek}, {DeRose}, {Cordero}, {Prat}, {Myles}, {Eckert}, {Bechtol}, {Herner}, {Secco}, {Gatti}, {Raveri}, {Kind}, {Becker}, {Troxel}, {Jarvis}, {MacCrann}, {Friedrich}, {Alves}, {Leget}, {Chen}, {Rollins}, {Wechsler}, {Gruendl}, {Cawthon}, {Allam}, {Bridle}, {Pandey}, {Everett}, {Shin}, {Hartley}, {Fang}, {Zhang}, {Aguena}, {Annis}, {Bacon}, {Bertin}, {Bocquet}, {Brooks}, {Carretero}, {Castander}, {da Costa}, {Pereira}, {De Vicente}, {Desai}, {Doel}, {Ferrero},
  {Flaugher}, {Frieman}, {Garc{\'\i}a-Bellido}, {Gaztanaga}, {Gutierrez}, {Hinton}, {Hollowood}, {Honscheid}, {Huterer}, {James}, {Kuehn}, {Lahav}, {Lee}, {Marshall}, {Mena-Fern{\'a}ndez}, {Miquel}, {Muir}, {Paterno}, {Plazas Malag{\'o}n}, {Porredon}, {Romer}, {Samuroff}, {Sanchez}, {Sanchez Cid}, {Smith}, {Soares-Santos}, {Suchyta}, {Swanson}, {Tarle}, {To}, {Weaverdyck}, {Weller}, {Wiseman}, \& {Yamamoto}}]{2024MNRAS.534..655B}
{Bigwood}, L., {Amon}, A., {Schneider}, A., {et~al.} 2024, \mnras, 534, 655, \dodoi{10.1093/mnras/stae2100}

\bibitem[{{Bleem} {et~al.}(2024){Bleem}, {Klein}, {Abbot}, {Ade}, {Aguena}, {Alves}, {Anderson}, {Andrade-Oliveira}, {Ansarinejad}, {Archipley}, {Ashby}, {Austermann}, {Bacon}, {Beall}, {Bender}, {Benson}, {Bianchini}, {Bocquet}, {Brooks}, {Burke}, {Calzadilla}, {Carlstrom}, {Carnero Rosell}, {Carretero}, {Chang}, {Chaubal}, {Chiang}, {Chou}, {Citron}, {Corbett Moran}, {Costanzi}, {Constanzi}, {Crawford}, {Crites}, {da Costa}, {de Haan}, {De Vicente}, {Desai}, {Dobbs}, {Doel}, {Everett}, {Ferrero}, {Flaugher}, {Floyd}, {Friedel}, {Frieman}, {Gallicchio}, {Garc'ia-Bellido}, {Gatti}, {George}, {Giannini}, {Grandis}, {Gruen}, {Gruendl}, {Gupta}, {Gutierrez}, {Halverson}, {Hinton}, {Hinton}, {Holder}, {Hollowood}, {Holzapfel}, {Honscheid}, {Hrubes}, {Huang}, {Hubmayr}, {Irwin}, {Mena-Fern{\'a}ndez}, {James}, {K{\'e}ruzor{\'e}}, {Knox}, {Kuehn}, {Lahav}, {Lee}, {Lee}, {Li}, {Lowitz}, {Marshal}, {McDonald}, {McMahon}, {Menanteau}, {Meyer}, {Miquel}, {Mohr}, {Montgomery}, {Myles}, {Natoli}, {Nibarger}, {Noble},
  {Novosad}, {Ogando}, {Padin}, {Patil}, {Pereira}, {Pieres}, {Plazas Malag'on}, {Pryke}, {Reichardt}, {Rodr'iguez-Monroy}, {Romer}, {Ruhl}, {Saliwanchik}, {Salvati}, {Sanchez}, {Saro}, {Schaffer}, {Schrabback}, {Sevilla-Noarbe}, {Sievers}, {Smecher}, {Smith}, {Somboonpanyakul}, {Stalder}, {Stark}, {Suchyta}, {Swanson}, {Tarle}, {To}, {Tucker}, {Veach}, {Vieira}, {Vincenzi}, {Wang}, {Weller}, {Whitehorn}, {Wiseman}, {Wu}, {Yefremenko}, {Zebrowski}, \& {Zhang}}]{2024OJAp....7E..13B}
{Bleem}, L.~E., {Klein}, M., {Abbot}, T.~M.~C., {et~al.} 2024, The Open Journal of Astrophysics, 7, 13, \dodoi{10.21105/astro.2311.07512}

\bibitem[{{Calafut} {et~al.}(2021){Calafut}, {Gallardo}, {Vavagiakis}, {Amodeo}, {Aiola}, {Austermann}, {Battaglia}, {Battistelli}, {Beall}, {Bean}, {Bond}, {Calabrese}, {Choi}, {Cothard}, {Devlin}, {Duell}, {Duff}, {Duivenvoorden}, {Dunkley}, {Dunner}, {Ferraro}, {Guan}, {Hill}, {Hilton}, {Hilton}, {Hlo{\v{z}}ek}, {Huber}, {Hubmayr}, {Huffenberger}, {Hughes}, {Koopman}, {Kosowsky}, {Li}, {Lokken}, {Madhavacheril}, {McMahon}, {Moodley}, {Naess}, {Nati}, {Newburgh}, {Niemack}, {Page}, {Partridge}, {Schaan}, {Schillaci}, {Sif{\'o}n}, {Spergel}, {Staggs}, {Ullom}, {Vale}, {Van Engelen}, {Van Lanen}, {Wollack}, \& {Xu}}]{2021PhRvD.104d3502C}
{Calafut}, V., {Gallardo}, P.~A., {Vavagiakis}, E.~M., {et~al.} 2021, \prd, 104, 043502, \dodoi{10.1103/PhysRevD.104.043502}

\bibitem[{{Carilli} \& {Rawlings}(2004)}]{2004NewAR..48..979C}
{Carilli}, C.~L., \& {Rawlings}, S. 2004, \nar, 48, 979, \dodoi{10.1016/j.newar.2004.09.001}

\bibitem[{{CHIME/FRB Collaboration} {et~al.}(2025){CHIME/FRB Collaboration}, {Amiri}, {Andersen}, {Andrew}, {Bandura}, {Bhardwaj}, {Bhopi}, {Bidula}, {Boyle}, {Brar}, {Carlson}, {Cassanelli}, {Cassity}, {Chatterjee}, {Cliche}, {Curtin}, {Darlinger}, {DeBoer}, {Dobbs}, {Dong}, {Eadie}, {Fonseca}, {Gaensler}, {Gusinskaia}, {Halpern}, {Hendricksen}, {Hessels}, {Joseph}, {Kaczmarek}, {Kaspi}, {Khairy}, {Landecker}, {Lanman}, {Kit Lau}, {Lazda}, {Leung}, {Main}, {Masui}, {Mckinven}, {Mena-Parra}, {Meyers}, {Michilli}, {Milutinovic}, {Nimmo}, {Noble}, {Pandhi}, {Pearlman}, {Peterson}, {Petroff}, {Pleunis}, {Pollak}, {Rafiei-Ravandi}, {Renard}, {Sammons}, {Sand}, {Sanghavi}, {Scholz}, {Shah}, {Shin}, {Siegel}, {Siemion}, {Sievers}, {Smith}, {Spear}, {Stairs}, {Vanderlinde}, {Wang}, {Willis}, \& {Zegmott}}]{2025arXiv250405192F}
{CHIME/FRB Collaboration}, {Amiri}, M., {Andersen}, B.~C., {et~al.} 2025, arXiv e-prints, arXiv:2504.05192, \dodoi{10.48550/arXiv.2504.05192}

\bibitem[{{Chisari} {et~al.}(2019){Chisari}, {Mead}, {Joudaki}, {Ferreira}, {Schneider}, {Mohr}, {Tr{\"o}ster}, {Alonso}, {McCarthy}, {Martin-Alvarez}, {Devriendt}, {Slyz}, \& {van Daalen}}]{2019OJAp....2E...4C}
{Chisari}, N.~E., {Mead}, A.~J., {Joudaki}, S., {et~al.} 2019, The Open Journal of Astrophysics, 2, 4, \dodoi{10.21105/astro.1905.06082}

\bibitem[{{Connor} {et~al.}(2024){Connor}, {Ravi}, {Sharma}, {Ocker}, {Faber}, {Hallinan}, {Harnach}, {Hellbourg}, {Hobbs}, {Hodge}, {Hodges}, {Kosogorov}, {Lamb}, {Law}, {Rasmussen}, {Sherman}, {Somalwar}, {Weinreb}, \& {Woody}}]{2024arXiv240916952C}
{Connor}, L., {Ravi}, V., {Sharma}, K., {et~al.} 2024, arXiv e-prints, arXiv:2409.16952, \dodoi{10.48550/arXiv.2409.16952}

\bibitem[{Cordes \& {Lazio}(2002)}]{2002astro.ph..7156C}
Cordes, J.~M., \& {Lazio}, T.~J.~W. 2002, arXiv e-prints, astro, \dodoi{10.48550/arXiv.astro-ph/0207156}

\bibitem[{{Cordes} \& {Lazio}(2003)}]{2003astro.ph..1598C}
{Cordes}, J.~M., \& {Lazio}, T.~J.~W. 2003, arXiv e-prints, astro, \dodoi{10.48550/arXiv.astro-ph/0301598}

\bibitem[{{Crill} {et~al.}(2020){Crill}, {Werner}, {Akeson}, {Ashby}, {Bleem}, {Bock}, {Bryan}, {Burnham}, {Byunh}, {Chang}, {Chiang}, {Cook}, {Cooray}, {Davis}, {Dor{\'e}}, {Dowell}, {Dubois-Felsmann}, {Eifler}, {Faisst}, {Habib}, {Heinrich}, {Heitmann}, {Heaton}, {Hirata}, {Hristov}, {Hui}, {Jeong}, {Kang}, {Kecman}, {Kirkpatrick}, {Korngut}, {Krause}, {Lee}, {Lisse}, {Masters}, {Mauskopf}, {Melnick}, {Miyasaka}, {Nayyeri}, {Nguyen}, {{\"O}berg}, {Padin}, {Paladini}, {Pourrahmani}, {Pyo}, {Smith}, {Song}, {Symons}, {Teplitz}, {Tolls}, {Unwin}, {Windhorst}, {Yang}, \& {Zemcov}}]{2020SPIE11443E..0IC}
{Crill}, B.~P., {Werner}, M., {Akeson}, R., {et~al.} 2020, in Society of Photo-Optical Instrumentation Engineers (SPIE) Conference Series, Vol. 11443, Space Telescopes and Instrumentation 2020: Optical, Infrared, and Millimeter Wave, ed. M.~{Lystrup} \& M.~D. {Perrin}, 114430I, \dodoi{10.1117/12.2567224}

\bibitem[{{Dalal} {et~al.}(2025){Dalal}, {To}, {Hirata}, {Hyeon-Shin}, {Hilton}, {Pandey}, \& {Bond}}]{2025arXiv250704476D}
{Dalal}, N., {To}, C.-H., {Hirata}, C., {et~al.} 2025, arXiv e-prints, arXiv:2507.04476, \dodoi{10.48550/arXiv.2507.04476}

\bibitem[{{DESI Collaboration} {et~al.}(2022){DESI Collaboration}, {Abareshi}, {Aguilar}, {Ahlen}, {Alam}, {Alexander}, {Alfarsy}, {Allen}, {Allende Prieto}, {Alves}, {Ameel}, {Armengaud}, {Asorey}, {Aviles}, {Bailey}, {Balaguera-Antol{\'\i}nez}, {Ballester}, {Baltay}, {Bault}, {Beltran}, {Benavides}, {BenZvi}, {Berti}, {Besuner}, {Beutler}, {Bianchi}, {Blake}, {Blanc}, {Blum}, {Bolton}, {Bose}, {Bramall}, {Brieden}, {Brodzeller}, {Brooks}, {Brownewell}, {Buckley-Geer}, {Cahn}, {Cai}, {Canning}, {Capasso}, {Carnero Rosell}, {Carton}, {Casas}, {Castander}, {Cervantes-Cota}, {Chabanier}, {Chaussidon}, {Chuang}, {Circosta}, {Cole}, {Cooper}, {da Costa}, {Cousinou}, {Cuceu}, {Davis}, {Dawson}, {de la Cruz-Noriega}, {de la Macorra}, {de Mattia}, {Della Costa}, {Demmer}, {Derwent}, {Dey}, {Dey}, {Dhungana}, {Ding}, {Dobson}, {Doel}, {Donald-McCann}, {Donaldson}, {Douglass}, {Duan}, {Dunlop}, {Edelstein}, {Eftekharzadeh}, {Eisenstein}, {Enriquez-Vargas}, {Escoffier}, {Evatt}, {Fagrelius}, {Fan}, {Fanning},
  {Fawcett}, {Ferraro}, {Ereza}, {Flaugher}, {Font-Ribera}, {Forero-Romero}, {Frenk}, {Fromenteau}, {G{\"a}nsicke}, {Garcia-Quintero}, {Garrison}, {Gazta{\~n}aga}, {Gerardi}, {Gil-Mar{\'\i}n}, {Gontcho A Gontcho}, {Gonzalez-Morales}, {Gonzalez-de-Rivera}, {Gonzalez-Perez}, {Gordon}, {Graur}, {Green}, {Grove}, {Gruen}, {Gutierrez}, {Guy}, {Hahn}, {Harris}, {Herrera}, {Herrera-Alcantar}, {Honscheid}, {Howlett}, {Huterer}, {Ir{\v{s}}i{\v{c}}}, {Ishak}, {Jelinsky}, {Jiang}, {Jimenez}, {Jing}, {Joyce}, {Jullo}, {Juneau}, {Kara{\c{c}}ayl{\i}}, {Karamanis}, {Karcher}, {Karim}, {Kehoe}, {Kent}, {Kirkby}, {Kisner}, {Kitaura}, {Koposov}, {Kov{\'a}cs}, {Kremin}, {Krolewski}, {L'Huillier}, {Lahav}, {Lambert}, {Lamman}, {Lan}, {Landriau}, {Lane}, {Lang}, {Lange}, {Lasker}, {Le Guillou}, {Leauthaud}, {Le Van Suu}, {Levi}, {Li}, {Magneville}, {Manera}, {Manser}, {Marshall}, {Martini}, {McCollam}, {McDonald}, {Meisner}, {Mena-Fern{\'a}ndez}, {Meneses-Rizo}, {Mezcua}, {Miller}, {Miquel}, {Montero-Camacho}, {Moon},
  {Moustakas}, {Mueller}, {Mu{\~n}oz-Guti{\'e}rrez}, {Myers}, {Nadathur}, {Najita}, {Napolitano}, {Neilsen}, {Newman}, {Nie}, {Ning}, {Niz}, {Norberg}, {Noriega}, {O'Brien}, {Obuljen}, {Palanque-Delabrouille}, {Palmese}, {Zhiwei}, {Pappalardo}, {PENG}, {Percival}, {Perruchot}, {Pogge}, {Poppett}, {Porredon}, {Prada}, {Prochaska}, {Pucha}, {P{\'e}rez-Fern{\'a}ndez}, {P{\'e}rez-R{\`a}fols}, {Rabinowitz}, \& {Raichoor}}]{2022AJ....164..207D}
{DESI Collaboration}, {Abareshi}, B., {Aguilar}, J., {et~al.} 2022, \aj, 164, 207, \dodoi{10.3847/1538-3881/ac882b}

\bibitem[{{DESI Collaboration} {et~al.}(2024){DESI Collaboration}, {Adame}, {Aguilar}, {Ahlen}, {Alam}, {Aldering}, {Alexander}, {Alfarsy}, {Allende Prieto}, {Alvarez}, {Alves}, {Anand}, {Andrade-Oliveira}, {Armengaud}, {Asorey}, {Avila}, {Aviles}, {Bailey}, {Balaguera-Antol{\'\i}nez}, {Ballester}, {Baltay}, {Bault}, {Bautista}, {Behera}, {Beltran}, {BenZvi}, {Beraldo e Silva}, {Bermejo-Climent}, {Berti}, {Besuner}, {Beutler}, {Bianchi}, {Blake}, {Blum}, {Bolton}, {Brieden}, {Brodzeller}, {Brooks}, {Brown}, {Buckley-Geer}, {Burtin}, {Cabayol-Garcia}, {Cai}, {Canning}, {Cardiel-Sas}, {Carnero Rosell}, {Castander}, {Cervantes-Cota}, {Chabanier}, {Chaussidon}, {Chaves-Montero}, {Chen}, {Chen}, {Chuang}, {Claybaugh}, {Cole}, {Cooper}, {Cuceu}, {Davis}, {Dawson}, {de Belsunce}, {de la Cruz}, {de la Macorra}, {Della Costa}, {de Mattia}, {Demina}, {Demirbozan}, {DeRose}, {Dey}, {Dey}, {Dhungana}, {Ding}, {Ding}, {Doel}, {Doshi}, {Douglass}, {Edge}, {Eftekharzadeh}, {Eisenstein}, {Elliott}, {Ereza}, {Escoffier},
  {Fagrelius}, {Fan}, {Fanning}, {Fawcett}, {Ferraro}, {Flaugher}, {Font-Ribera}, {Forero-Romero}, {Forero-S{\'a}nchez}, {Frenk}, {G{\"a}nsicke}, {Garc{\'\i}a}, {Garc{\'\i}a-Bellido}, {Garcia-Quintero}, {Garrison}, {Gil-Mar{\'\i}n}, {Golden-Marx}, {Gontcho A Gontcho}, {Gonzalez-Morales}, {Gonzalez-Perez}, {Gordon}, {Graur}, {Green}, {Gruen}, {Guy}, {Hadzhiyska}, {Hahn}, {Han}, {Hanif}, {Herrera-Alcantar}, {Honscheid}, {Hou}, {Howlett}, {Huterer}, {Ir{\v{s}}i{\v{c}}}, {Ishak}, {Jacques}, {Jana}, {Jiang}, {Jimenez}, {Jing}, {Joudaki}, {Joyce}, {Jullo}, {Juneau}, {Kara{\c{c}}ayl{\i}}, {Karim}, {Kehoe}, {Kent}, {Khederlarian}, {Kim}, {Kirkby}, {Kisner}, {Kitaura}, {Kizhuprakkat}, {Kneib}, {Koposov}, {Kov{\'a}cs}, {Kremin}, {Krolewski}, {L'Huillier}, {Lahav}, {Lambert}, {Lamman}, {Lan}, {Landriau}, {Lang}, {Lange}, {Lasker}, {Leauthaud}, {Le Guillou}, {Levi}, {Li}, {Linder}, {Lyons}, {Magneville}, {Manera}, {Manser}, {Margala}, {Martini}, {McDonald}, {Medina}, {Medina-Varela}, {Meisner}, {Mena-Fern{\'a}ndez},
  {Meneses-Rizo}, {Mezcua}, {Miquel}, {Montero-Camacho}, {Moon}, {Moore}, {Moustakas}, {Mueller}, {Mundet}, {Mu{\~n}oz-Guti{\'e}rrez}, {Myers}, {Nadathur}, {Napolitano}, {Neveux}, {Newman}, {Nie}, {Nikutta}, {Niz}, {Norberg}, {Noriega}, {Paillas}, {Palanque-Delabrouille}, {Palmese}, {Pan}, {Parkinson}, {Penmetsa}, {Percival}, {P{\'e}rez-Fern{\'a}ndez}, {P{\'e}rez-R{\`a}fols}, {Pieri}, {Poppett}, {Porredon}, \& {Pothier}}]{2024AJ....168...58D}
{DESI Collaboration}, {Adame}, A.~G., {Aguilar}, J., {et~al.} 2024, \aj, 168, 58, \dodoi{10.3847/1538-3881/ad3217}

\bibitem[{{Di Valentino} {et~al.}(2025){Di Valentino}, {Levi Said}, {Riess}, {Pollo}, {Poulin}, {G{\'o}mez-Valent}, {Weltman}, {Palmese}, {Huang}, {van de Bruck}, {Shekhar Saraf}, {Kuo}, {Uhlemann}, {Grand{\'o}n}, {Paz}, {Eckert}, {Teixeira}, {Saridakis}, {Colg{\'a}in}, {Beutler}, {Niedermann}, {Bajardi}, {Barenboim}, {Gubitosi}, {Musella}, {Banik}, {Szapudi}, {Singal}, {Haro Cases}, {Chluba}, {Torrado}, {Mifsud}, {Jedamzik}, {Said}, {Dialektopoulos}, {Herold}, {Perivolaropoulos}, {Zu}, {Galbany}, {Breuval}, {Visinelli}, {Escamilla}, {Anchordoqui}, {Sheikh-Jabbari}, {Lembo}, {Dainotti}, {Vincenzi}, {Asgari}, {Gerbino}, {Forconi}, {Cantiello}, {Moresco}, {Benetti}, {Sch{\"o}neberg}, {Akarsu}, {Nunes}, {Bernardo}, {Ch{\'a}vez}, {Anderson}, {Watkins}, {Capozziello}, {Li}, {Vagnozzi}, {Pan}, {Treu}, {Irsic}, {Handley}, {Giar{\`e}}, {Murakami}, {Poudou}, {Heavens}, {Kogut}, {Domi}, {{\L}ukasz Lenart}, {Melchiorri}, {Vadal{\`a}}, {Amon}, {Bonilla}, {Reeves}, {Zhuk}, {Bonanno}, {{\"O}vg{\"u}n}, {Pisani}, {Talebian},
  {Abebe}, {Aboubrahim}, {Gonz{\'a}lez Mor{\'a}n}, {Kov{\'a}cs}, {Papatriantafyllou}, {Liddle}, {Paliathanasis}, {Borowiec}, {Yadav}, {Yadav}, {Sen}, {Mini Latha}, {Davis}, {Shajib}, {Walters}, {Idicherian Lonappan}, {Chudaykin}, {Capodagli}, {da Silva}, {De Felice}, {Racioppi}, {Soler Oficial}, {Montiel}, {Favale}, {Bernui}, {Velasco}, {Heinesen}, {Bakopoulos}, {Chatzistavrakidis}, {Khanpour}, {Sathyaprakash}, {Zgirski}, {L'Huillier}, {Famaey}, {Jain}, {Marek}, {Zhang}, {Karmakar}, {Dragovich}, {Thomas}, {Correa}, {Boiza}, {Marques}, {Escamilla-Rivera}, {Tzerefos}, {Zhang}, {De Leo}, {Pfeifer}, {Lee}, {Venter}, {Gomes}, {Roque De bom}, {Moreno-Pulido}, {Iosifidis}, {Grin}, {Blixt}, {Scolnic}, {Oriti}, {Dobrycheva}, {Bettoni}, {Benisty}, {Fern{\'a}ndez-Arenas}, {Wiltshire}, {Sanchez Cid}, {Tamayo}, {Valls-Gabaud}, {Pedrotti}, {Wang}, {Staicova}, {Totolou}, {Rubiera-Garcia}, {Milakovi{\'c}}, {Pesce}, {Sluse}, {Borka}, {Yusofi}, {Giusarma}, {Terlevich}, {Tomasetti}, {Vagenas}, {Fazzari}, {Ferreira},
  {Barakovic}, {Dimastrogiovanni}, {Brinch Holm}, {Mottola}, {{\"O}z{\"u}lker}, {Specogna}, {Brocato}, {Jensko}, {Antonette Enriquez}, {Bhatia}, {Bresolin}, {Avila}, {Bouch{\`e}}, {Bombacigno}, {Anagnostopoulos}, {Pace}, {Sorrenti}, {Lobo}, {Courbin}, {Hansen}, {Sloan}, {Farrugia}, {Lynch}, {Garcia-Arroyo}, {Raimondo}, {Lambiase}, {Anand}, {Poulot}, {Leon}, {Kouniatalis}, {Nardini}, {Cs{\"o}rnyei}, {Galloni}, \& {Bargiacchi}}]{2025arXiv250401669D}
{Di Valentino}, E., {Levi Said}, J., {Riess}, A., {et~al.} 2025, arXiv e-prints, arXiv:2504.01669, \dodoi{10.48550/arXiv.2504.01669}

\bibitem[{{Euclid Collaboration} {et~al.}(2020){Euclid Collaboration}, {Blanchard}, {Camera}, {Carbone}, {Cardone}, {Casas}, {Clesse}, {Ili{\'c}}, {Kilbinger}, {Kitching}, {Kunz}, {Lacasa}, {Linder}, {Majerotto}, {Markovi{\v{c}}}, {Martinelli}, {Pettorino}, {Pourtsidou}, {Sakr}, {S{\'a}nchez}, {Sapone}, {Tutusaus}, {Yahia-Cherif}, {Yankelevich}, {Andreon}, {Aussel}, {Balaguera-Antol{\'\i}nez}, {Baldi}, {Bardelli}, {Bender}, {Biviano}, {Bonino}, {Boucaud}, {Bozzo}, {Branchini}, {Brau-Nogue}, {Brescia}, {Brinchmann}, {Burigana}, {Cabanac}, {Capobianco}, {Cappi}, {Carretero}, {Carvalho}, {Casas}, {Castander}, {Castellano}, {Cavuoti}, {Cimatti}, {Cledassou}, {Colodro-Conde}, {Congedo}, {Conselice}, {Conversi}, {Copin}, {Corcione}, {Coupon}, {Courtois}, {Cropper}, {Da Silva}, {de la Torre}, {Di Ferdinando}, {Dubath}, {Ducret}, {Duncan}, {Dupac}, {Dusini}, {Fabbian}, {Fabricius}, {Farrens}, {Fosalba}, {Fotopoulou}, {Fourmanoit}, {Frailis}, {Franceschi}, {Franzetti}, {Fumana}, {Galeotta}, {Gillard}, {Gillis},
  {Giocoli}, {G{\'o}mez-Alvarez}, {Graci{\'a}-Carpio}, {Grupp}, {Guzzo}, {Hoekstra}, {Hormuth}, {Israel}, {Jahnke}, {Keihanen}, {Kermiche}, {Kirkpatrick}, {Kohley}, {Kubik}, {Kurki-Suonio}, {Ligori}, {Lilje}, {Lloro}, {Maino}, {Maiorano}, {Marggraf}, {Martinet}, {Marulli}, {Massey}, {Medinaceli}, {Mei}, {Mellier}, {Metcalf}, {Metge}, {Meylan}, {Moresco}, {Moscardini}, {Munari}, {Nichol}, {Niemi}, {Nucita}, {Padilla}, {Paltani}, {Pasian}, {Percival}, {Pires}, {Polenta}, {Poncet}, {Pozzetti}, {Racca}, {Raison}, {Renzi}, {Rhodes}, {Romelli}, {Roncarelli}, {Rossetti}, {Saglia}, {Schneider}, {Scottez}, {Secroun}, {Sirri}, {Stanco}, {Starck}, {Sureau}, {Tallada-Cresp{\'\i}}, {Tavagnacco}, {Taylor}, {Tenti}, {Tereno}, {Toledo-Moreo}, {Torradeflot}, {Valenziano}, {Vassallo}, {Verdoes Kleijn}, {Viel}, {Wang}, {Zacchei}, {Zoubian}, \& {Zucca}}]{2020A&A...642A.191E}
{Euclid Collaboration}, {Blanchard}, A., {Camera}, S., {et~al.} 2020, \aap, 642, A191, \dodoi{10.1051/0004-6361/202038071}

\bibitem[{{Euclid Collaboration} {et~al.}(2022){Euclid Collaboration}, {Scaramella}, {Amiaux}, {Mellier}, {Burigana}, {Carvalho}, {Cuillandre}, {Da Silva}, {Derosa}, {Dinis}, {Maiorano}, {Maris}, {Tereno}, {Laureijs}, {Boenke}, {Buenadicha}, {Dupac}, {Gaspar Venancio}, {G{\'o}mez-{\'A}lvarez}, {Hoar}, {Lorenzo Alvarez}, {Racca}, {Saavedra-Criado}, {Schwartz}, {Vavrek}, {Schirmer}, {Aussel}, {Azzollini}, {Cardone}, {Cropper}, {Ealet}, {Garilli}, {Gillard}, {Granett}, {Guzzo}, {Hoekstra}, {Jahnke}, {Kitching}, {Maciaszek}, {Meneghetti}, {Miller}, {Nakajima}, {Niemi}, {Pasian}, {Percival}, {Pottinger}, {Sauvage}, {Scodeggio}, {Wachter}, {Zacchei}, {Aghanim}, {Amara}, {Auphan}, {Auricchio}, {Awan}, {Balestra}, {Bender}, {Bodendorf}, {Bonino}, {Branchini}, {Brau-Nogue}, {Brescia}, {Candini}, {Capobianco}, {Carbone}, {Carlberg}, {Carretero}, {Casas}, {Castander}, {Castellano}, {Cavuoti}, {Cimatti}, {Cledassou}, {Congedo}, {Conselice}, {Conversi}, {Copin}, {Corcione}, {Costille}, {Courbin}, {Degaudenzi}, {Douspis},
  {Dubath}, {Duncan}, {Dusini}, {Farrens}, {Ferriol}, {Fosalba}, {Fourmanoit}, {Frailis}, {Franceschi}, {Franzetti}, {Fumana}, {Gillis}, {Giocoli}, {Grazian}, {Grupp}, {Haugan}, {Holmes}, {Hormuth}, {Hudelot}, {Kermiche}, {Kiessling}, {Kilbinger}, {Kohley}, {Kubik}, {K{\"u}mmel}, {Kunz}, {Kurki-Suonio}, {Lahav}, {Ligori}, {Lilje}, {Lloro}, {Mansutti}, {Marggraf}, {Markovic}, {Marulli}, {Massey}, {Maurogordato}, {Melchior}, {Merlin}, {Meylan}, {Mohr}, {Moresco}, {Morin}, {Moscardini}, {Munari}, {Nichol}, {Padilla}, {Paltani}, {Peacock}, {Pedersen}, {Pettorino}, {Pires}, {Poncet}, {Popa}, {Pozzetti}, {Raison}, {Rebolo}, {Rhodes}, {Rix}, {Roncarelli}, {Rossetti}, {Saglia}, {Schneider}, {Schrabback}, {Secroun}, {Seidel}, {Serrano}, {Sirignano}, {Sirri}, {Skottfelt}, {Stanco}, {Starck}, {Tallada-Cresp{\'\i}}, {Tavagnacco}, {Taylor}, {Teplitz}, {Toledo-Moreo}, {Torradeflot}, {Trifoglio}, {Valentijn}, {Valenziano}, {Verdoes Kleijn}, {Wang}, {Welikala}, {Weller}, {Wetzstein}, {Zamorani}, {Zoubian}, {Andreon},
  {Baldi}, {Bardelli}, {Boucaud}, {Camera}, {Di Ferdinando}, {Fabbian}, {Farinelli}, {Galeotta}, {Graci{\'a}-Carpio}, {Maino}, {Medinaceli}, {Mei}, {Neissner}, {Polenta}, {Renzi}, {Romelli}, {Rosset}, {Sureau}, {Tenti}, {Vassallo}, {Zucca}, {Baccigalupi}, {Balaguera-Antol{\'\i}nez}, {Battaglia}, {Biviano}, {Borgani}, {Bozzo}, {Cabanac}, \& {Cappi}}]{2022A&A...662A.112E}
{Euclid Collaboration}, {Scaramella}, R., {Amiaux}, J., {et~al.} 2022, \aap, 662, A112, \dodoi{10.1051/0004-6361/202141938}

\bibitem[{{Ferreira} {et~al.}(2024){Ferreira}, {Alonso}, {Garcia-Garcia}, \& {Chisari}}]{2024PhRvL.133e1001F}
{Ferreira}, T., {Alonso}, D., {Garcia-Garcia}, C., \& {Chisari}, N.~E. 2024, \prl, 133, 051001, \dodoi{10.1103/PhysRevLett.133.051001}

\bibitem[{{Foley} {et~al.}(2018){Foley}, {Koekemoer}, {Spergel}, {Bianco}, {Capak}, {Dai}, {Dore}, {Fazio}, {Ferguson}, {Filippenko}, {Frye}, {Galbany}, {Gawiser}, {Gronwall}, {Hathi}, {Hirata}, {Hounsell}, {Jha}, {Kim}, {Kelly}, {Kruk}, {Malhotra}, {Mandel}, {Margutti}, {Marrone}, {McQuinn}, {Melchior}, {Moustakas}, {Newman}, {Peek}, {Perlmutter}, {Rhodes}, {Robertson}, {Rubin}, {Scolnic}, {Somerville}, {Street}, {Wang}, {Whalen}, {Windhorst}, \& {Wollack}}]{2018arXiv181200514F}
{Foley}, R.~J., {Koekemoer}, A.~M., {Spergel}, D.~N., {et~al.} 2018, arXiv e-prints, arXiv:1812.00514, \dodoi{10.48550/arXiv.1812.00514}

\bibitem[{{Ghirardini} {et~al.}(2024){Ghirardini}, {Bulbul}, {Artis}, {Clerc}, {Garrel}, {Grandis}, {Kluge}, {Liu}, {Bahar}, {Balzer}, {Chiu}, {Comparat}, {Gruen}, {Kleinebreil}, {Krippendorf}, {Merloni}, {Nandra}, {Okabe}, {Pacaud}, {Predehl}, {Ramos-Ceja}, {Reiprich}, {Sanders}, {Schrabback}, {Seppi}, {Zelmer}, {Zhang}, {Bornemann}, {Brunner}, {Burwitz}, {Coutinho}, {Dennerl}, {Freyberg}, {Friedrich}, {Gaida}, {Gueguen}, {Haberl}, {Kink}, {Lamer}, {Li}, {Liu}, {Maitra}, {Meidinger}, {Mueller}, {Miyatake}, {Miyazaki}, {Robrade}, {Schwope}, \& {Stewart}}]{2024A&A...689A.298G}
{Ghirardini}, V., {Bulbul}, E., {Artis}, E., {et~al.} 2024, \aap, 689, A298, \dodoi{10.1051/0004-6361/202348852}

\bibitem[{{Grandis} {et~al.}(2024){Grandis}, {Aric{\`o}}, {Schneider}, \& {Linke}}]{2024MNRAS.528.4379G}
{Grandis}, S., {Aric{\`o}}, G., {Schneider}, A., \& {Linke}, L. 2024, \mnras, 528, 4379, \dodoi{10.1093/mnras/stae259}

\bibitem[{{Hadzhiyska} {et~al.}(2025){Hadzhiyska}, {Ferraro}, {Farren}, {Sailer}, \& {Zhou}}]{2025arXiv250714136H}
{Hadzhiyska}, B., {Ferraro}, S., {Farren}, G.~S., {Sailer}, N., \& {Zhou}, R. 2025, arXiv e-prints, arXiv:2507.14136, \dodoi{10.48550/arXiv.2507.14136}

\bibitem[{{Hadzhiyska} {et~al.}(2024){Hadzhiyska}, {Ferraro}, {Ried Guachalla}, {Schaan}, {Aguilar}, {Battaglia}, {Bond}, {Brooks}, {Calabrese}, {Choi}, {Claybaugh}, {Coulton}, {Dawson}, {Devlin}, {Dey}, {Doel}, {Duivenvoorden}, {Dunkley}, {Farren}, {Font-Ribera}, {Forero-Romero}, {Gallardo}, {Gazta{\~n}aga}, {Gontcho Gontcho}, {Gralla}, {Le Guillou}, {Gutierrez}, {Guy}, {Hill}, {Hlo{\v{z}}ek}, {Honscheid}, {Juneau}, {Kisner}, {Kremin}, {Landriau}, {Liu}, {Louis}, {MacCrann}, {de Macorra}, {Madhavacheril}, {Manera}, {Meisner}, {Miquel}, {Moodley}, {Moustakas}, {Mroczkowski}, {Naess}, {Newman}, {Niemack}, {Niz}, {Page}, {Palanque-Delabrouille}, {Partridge}, {Percival}, {Prada}, {Qu}, {Rossi}, {Sanchez}, {Schlegel}, {Schubnell}, {Sehgal}, {Seo}, {Sif{\'o}n}, {Spergel}, {Sprayberry}, {Staggs}, {Tarl{\'e}}, {Vargas}, {Vavagiakis}, {Weaver}, {Wollack}, {Zhou}, \& {Zou}}]{2024arXiv240707152H}
{Hadzhiyska}, B., {Ferraro}, S., {Ried Guachalla}, B., {et~al.} 2024, arXiv e-prints, arXiv:2407.07152, \dodoi{10.48550/arXiv.2407.07152}

\bibitem[{{Hagstotz} {et~al.}(2022){Hagstotz}, {Reischke}, \& {Lilow}}]{2022MNRAS.511..662H}
{Hagstotz}, S., {Reischke}, R., \& {Lilow}, R. 2022, \mnras, 511, 662, \dodoi{10.1093/mnras/stac077}

\bibitem[{{Hallinan} {et~al.}(2019){Hallinan}, {Ravi}, {Weinreb}, {Kocz}, {Huang}, {Woody}, {Lamb}, {D'Addario}, {Catha}, {Law}, {Kulkarni}, {Phinney}, {Eastwood}, {Bouman}, {McLaughlin}, {Ransom}, {Siemens}, {Cordes}, {Lynch}, {Kaplan}, {Brazier}, {Bhatnagar}, {Myers}, {Walter}, \& {Gaensler}}]{2019BAAS...51g.255H}
{Hallinan}, G., {Ravi}, V., {Weinreb}, S., {et~al.} 2019, in Bulletin of the American Astronomical Society, Vol.~51, 255, \dodoi{10.48550/arXiv.1907.07648}

\bibitem[{{Hearin} {et~al.}(2012){Hearin}, {Zentner}, \& {Ma}}]{2012JCAP...04..034H}
{Hearin}, A.~P., {Zentner}, A.~R., \& {Ma}, Z. 2012, \jcap, 2012, 034, \dodoi{10.1088/1475-7516/2012/04/034}

\bibitem[{{Ivezi{\'c}} {et~al.}(2019){Ivezi{\'c}}, {Kahn}, {Tyson}, {Abel}, {Acosta}, {Allsman}, {Alonso}, {AlSayyad}, {Anderson}, {Andrew}, {Angel}, {Angeli}, {Ansari}, {Antilogus}, {Araujo}, {Armstrong}, {Arndt}, {Astier}, {Aubourg}, {Auza}, {Axelrod}, {Bard}, {Barr}, {Barrau}, {Bartlett}, {Bauer}, {Bauman}, {Baumont}, {Bechtol}, {Bechtol}, {Becker}, {Becla}, {Beldica}, {Bellavia}, {Bianco}, {Biswas}, {Blanc}, {Blazek}, {Blandford}, {Bloom}, {Bogart}, {Bond}, {Booth}, {Borgland}, {Borne}, {Bosch}, {Boutigny}, {Brackett}, {Bradshaw}, {Brandt}, {Brown}, {Bullock}, {Burchat}, {Burke}, {Cagnoli}, {Calabrese}, {Callahan}, {Callen}, {Carlin}, {Carlson}, {Chandrasekharan}, {Charles-Emerson}, {Chesley}, {Cheu}, {Chiang}, {Chiang}, {Chirino}, {Chow}, {Ciardi}, {Claver}, {Cohen-Tanugi}, {Cockrum}, {Coles}, {Connolly}, {Cook}, {Cooray}, {Covey}, {Cribbs}, {Cui}, {Cutri}, {Daly}, {Daniel}, {Daruich}, {Daubard}, {Daues}, {Dawson}, {Delgado}, {Dellapenna}, {de Peyster}, {de Val-Borro}, {Digel}, {Doherty}, {Dubois},
  {Dubois-Felsmann}, {Durech}, {Economou}, {Eifler}, {Eracleous}, {Emmons}, {Fausti Neto}, {Ferguson}, {Figueroa}, {Fisher-Levine}, {Focke}, {Foss}, {Frank}, {Freemon}, {Gangler}, {Gawiser}, {Geary}, {Gee}, {Geha}, {Gessner}, {Gibson}, {Gilmore}, {Glanzman}, {Glick}, {Goldina}, {Goldstein}, {Goodenow}, {Graham}, {Gressler}, {Gris}, {Guy}, {Guyonnet}, {Haller}, {Harris}, {Hascall}, {Haupt}, {Hernandez}, {Herrmann}, {Hileman}, {Hoblitt}, {Hodgson}, {Hogan}, {Howard}, {Huang}, {Huffer}, {Ingraham}, {Innes}, {Jacoby}, {Jain}, {Jammes}, {Jee}, {Jenness}, {Jernigan}, {Jevremovi{\'c}}, {Johns}, {Johnson}, {Johnson}, {Jones}, {Juramy-Gilles}, {Juri{\'c}}, {Kalirai}, {Kallivayalil}, {Kalmbach}, {Kantor}, {Karst}, {Kasliwal}, {Kelly}, {Kessler}, {Kinnison}, {Kirkby}, {Knox}, {Kotov}, {Krabbendam}, {Krughoff}, {Kub{\'a}nek}, {Kuczewski}, {Kulkarni}, {Ku}, {Kurita}, {Lage}, {Lambert}, {Lange}, {Langton}, {Le Guillou}, {Levine}, {Liang}, {Lim}, {Lintott}, {Long}, {Lopez}, {Lotz}, {Lupton}, {Lust}, {MacArthur}, {Mahabal},
  {Mandelbaum}, {Markiewicz}, {Marsh}, {Marshall}, {Marshall}, {May}, {McKercher}, {McQueen}, {Meyers}, {Migliore}, {Miller}, \& {Mills}}]{2019ApJ...873..111I}
{Ivezi{\'c}}, {\v{Z}}., {Kahn}, S.~M., {Tyson}, J.~A., {et~al.} 2019, \apj, 873, 111, \dodoi{10.3847/1538-4357/ab042c}

\bibitem[{{James} {et~al.}(2022{\natexlab{a}}){James}, {Prochaska}, {Macquart}, {North-Hickey}, {Bannister}, \& {Dunning}}]{2022MNRAS.509.4775J}
{James}, C.~W., {Prochaska}, J.~X., {Macquart}, J.~P., {et~al.} 2022{\natexlab{a}}, \mnras, 509, 4775, \dodoi{10.1093/mnras/stab3051}

\bibitem[{{James} {et~al.}(2022{\natexlab{b}}){James}, {Ghosh}, {Prochaska}, {Bannister}, {Bhandari}, {Day}, {Deller}, {Glowacki}, {Gordon}, {Heintz}, {Marnoch}, {Ryder}, {Scott}, {Shannon}, \& {Tejos}}]{2022MNRAS.516.4862J}
{James}, C.~W., {Ghosh}, E.~M., {Prochaska}, J.~X., {et~al.} 2022{\natexlab{b}}, \mnras, 516, 4862, \dodoi{10.1093/mnras/stac2524}

\bibitem[{{Khrykin} {et~al.}(2024){Khrykin}, {Ata}, {Lee}, {Simha}, {Huang}, {Prochaska}, {Tejos}, {Bannister}, {Cooke}, {Day}, {Deller}, {Glowacki}, {Gordon}, {James}, {Marnoch}, {Shannon}, {Zhang}, \& {Bernales-Cortes}}]{2024arXiv240200505K}
{Khrykin}, I.~S., {Ata}, M., {Lee}, K.-G., {et~al.} 2024, arXiv e-prints, arXiv:2402.00505, \dodoi{10.48550/arXiv.2402.00505}

\bibitem[{{Kova{\v{c}}} {et~al.}(2025){Kova{\v{c}}}, {Nicola}, {Bucko}, {Schneider}, {Reischke}, {Giri}, {Teyssier}, {Schaller}, \& {Schaye}}]{2025arXiv250707991K}
{Kova{\v{c}}}, M., {Nicola}, A., {Bucko}, J., {et~al.} 2025, arXiv e-prints, arXiv:2507.07991, \dodoi{10.48550/arXiv.2507.07991}

\bibitem[{{Krause} \& {Eifler}(2017)}]{2017MNRAS.470.2100K}
{Krause}, E., \& {Eifler}, T. 2017, \mnras, 470, 2100, \dodoi{10.1093/mnras/stx1261}

\bibitem[{{Lau} {et~al.}(2024{\natexlab{a}}){Lau}, {Bogdan}, {Zhou}, {Nagai}, \& {Cappelluti}}]{2024eas..conf.1910L}
{Lau}, E., {Bogdan}, A., {Zhou}, Y., {Nagai}, D., \& {Cappelluti}, N. 2024{\natexlab{a}}, in EAS2024, European Astronomical Society Annual Meeting, 1910

\bibitem[{{Lau} {et~al.}(2024{\natexlab{b}}){Lau}, {Nagai}, {Bogdan}, {Oppenheimer}, {Genel}, {Villaescusa-Navarro}, \& {Angles-Alcazar}}]{2024eas..conf.1906L}
{Lau}, E., {Nagai}, D., {Bogdan}, A., {et~al.} 2024{\natexlab{b}}, in EAS2024, European Astronomical Society Annual Meeting, 1906

\bibitem[{{Lau} {et~al.}(2025){Lau}, {Bogd{\'a}n}, {Nagai}, {Cappelluti}, \& {Shirasaki}}]{2025ApJ...983....8L}
{Lau}, E.~T., {Bogd{\'a}n}, {\'A}., {Nagai}, D., {Cappelluti}, N., \& {Shirasaki}, M. 2025, \apj, 983, 8, \dodoi{10.3847/1538-4357/adba5b}

\bibitem[{{Laureijs} {et~al.}(2011){Laureijs}, {Amiaux}, {Arduini}, {Augu{\`e}res}, {Brinchmann}, {Cole}, {Cropper}, {Dabin}, {Duvet}, {Ealet}, {Garilli}, {Gondoin}, {Guzzo}, {Hoar}, {Hoekstra}, {Holmes}, {Kitching}, {Maciaszek}, {Mellier}, {Pasian}, {Percival}, {Rhodes}, {Saavedra Criado}, {Sauvage}, {Scaramella}, {Valenziano}, {Warren}, {Bender}, {Castander}, {Cimatti}, {Le F{\`e}vre}, {Kurki-Suonio}, {Levi}, {Lilje}, {Meylan}, {Nichol}, {Pedersen}, {Popa}, {Rebolo Lopez}, {Rix}, {Rottgering}, {Zeilinger}, {Grupp}, {Hudelot}, {Massey}, {Meneghetti}, {Miller}, {Paltani}, {Paulin-Henriksson}, {Pires}, {Saxton}, {Schrabback}, {Seidel}, {Walsh}, {Aghanim}, {Amendola}, {Bartlett}, {Baccigalupi}, {Beaulieu}, {Benabed}, {Cuby}, {Elbaz}, {Fosalba}, {Gavazzi}, {Helmi}, {Hook}, {Irwin}, {Kneib}, {Kunz}, {Mannucci}, {Moscardini}, {Tao}, {Teyssier}, {Weller}, {Zamorani}, {Zapatero Osorio}, {Boulade}, {Foumond}, {Di Giorgio}, {Guttridge}, {James}, {Kemp}, {Martignac}, {Spencer}, {Walton}, {Bl{\"u}mchen}, {Bonoli},
  {Bortoletto}, {Cerna}, {Corcione}, {Fabron}, {Jahnke}, {Ligori}, {Madrid}, {Martin}, {Morgante}, {Pamplona}, {Prieto}, {Riva}, {Toledo}, {Trifoglio}, {Zerbi}, {Abdalla}, {Douspis}, {Grenet}, {Borgani}, {Bouwens}, {Courbin}, {Delouis}, {Dubath}, {Fontana}, {Frailis}, {Grazian}, {Koppenh{\"o}fer}, {Mansutti}, {Melchior}, {Mignoli}, {Mohr}, {Neissner}, {Noddle}, {Poncet}, {Scodeggio}, {Serrano}, {Shane}, {Starck}, {Surace}, {Taylor}, {Verdoes-Kleijn}, {Vuerli}, {Williams}, {Zacchei}, {Altieri}, {Escudero Sanz}, {Kohley}, {Oosterbroek}, {Astier}, {Bacon}, {Bardelli}, {Baugh}, {Bellagamba}, {Benoist}, {Bianchi}, {Biviano}, {Branchini}, {Carbone}, {Cardone}, {Clements}, {Colombi}, {Conselice}, {Cresci}, {Deacon}, {Dunlop}, {Fedeli}, {Fontanot}, {Franzetti}, {Giocoli}, {Garcia-Bellido}, {Gow}, {Heavens}, {Hewett}, {Heymans}, {Holland}, {Huang}, {Ilbert}, {Joachimi}, {Jennins}, {Kerins}, {Kiessling}, {Kirk}, {Kotak}, {Krause}, {Lahav}, {van Leeuwen}, {Lesgourgues}, {Lombardi}, {Magliocchetti}, {Maguire},
  {Majerotto}, {Maoli}, {Marulli}, {Maurogordato}, {McCracken}, {McLure}, {Melchiorri}, {Merson}, {Moresco}, {Nonino}, {Norberg}, {Peacock}, {Pello}, {Penny}, {Pettorino}, {Di Porto}, {Pozzetti}, {Quercellini}, {Radovich}, {Rassat}, {Roche}, {Ronayette}, \& {Rossetti}}]{2011arXiv1110.3193L}
{Laureijs}, R., {Amiaux}, J., {Arduini}, S., {et~al.} 2011, arXiv e-prints, arXiv:1110.3193, \dodoi{10.48550/arXiv.1110.3193}

\bibitem[{{Lorimer} {et~al.}(2007){Lorimer}, {Bailes}, {McLaughlin}, {Narkevic}, \& {Crawford}}]{2007Sci...318..777L}
{Lorimer}, D.~R., {Bailes}, M., {McLaughlin}, M.~A., {Narkevic}, D.~J., \& {Crawford}, F. 2007, Science, 318, 777, \dodoi{10.1126/science.1147532}

\bibitem[{{Luo} {et~al.}(2018){Luo}, {Lee}, {Lorimer}, \& {Zhang}}]{2018MNRAS.481.2320L}
{Luo}, R., {Lee}, K., {Lorimer}, D.~R., \& {Zhang}, B. 2018, \mnras, 481, 2320, \dodoi{10.1093/mnras/sty2364}

\bibitem[{{Macquart} {et~al.}(2020){Macquart}, {Prochaska}, {McQuinn}, {Bannister}, {Bhandari}, {Day}, {Deller}, {Ekers}, {James}, {Marnoch}, {Os{\l}owski}, {Phillips}, {Ryder}, {Scott}, {Shannon}, \& {Tejos}}]{2020Natur.581..391M}
{Macquart}, J.~P., {Prochaska}, J.~X., {McQuinn}, M., {et~al.} 2020, \nat, 581, 391, \dodoi{10.1038/s41586-020-2300-2}

\bibitem[{{McQuinn}(2014)}]{2014ApJ...780L..33M}
{McQuinn}, M. 2014, \apjl, 780, L33, \dodoi{10.1088/2041-8205/780/2/L33}

\bibitem[{{Mead} {et~al.}(2020){Mead}, {Tr{\"o}ster}, {Heymans}, {Van Waerbeke}, \& {McCarthy}}]{2020A&A...641A.130M}
{Mead}, A.~J., {Tr{\"o}ster}, T., {Heymans}, C., {Van Waerbeke}, L., \& {McCarthy}, I.~G. 2020, \aap, 641, A130, \dodoi{10.1051/0004-6361/202038308}

\bibitem[{{Merryfield} {et~al.}(2023){Merryfield}, {Tendulkar}, {Shin}, {Andersen}, {Josephy}, {Good}, {Dong}, {Masui}, {Lang}, {M{\"u}nchmeyer}, {Brar}, {Cassanelli}, {Dobbs}, {Fonseca}, {Kaspi}, {Mena-Parra}, {Pleunis}, {Rafiei-Ravandi}, {Sand}, {Scholz}, {Smith}, \& {Stairs}}]{2023AJ....165..152M}
{Merryfield}, M., {Tendulkar}, S.~P., {Shin}, K., {et~al.} 2023, \aj, 165, 152, \dodoi{10.3847/1538-3881/ac9ab5}

\bibitem[{{Pandey} {et~al.}(2025{\natexlab{a}}){Pandey}, {Salcido}, {To}, {Hill}, {Anbajagane}, {Baxter}, \& {McCarthy}}]{2025PhRvD.111d3529P}
{Pandey}, S., {Salcido}, J., {To}, C.-H., {et~al.} 2025{\natexlab{a}}, \prd, 111, 043529, \dodoi{10.1103/PhysRevD.111.043529}

\bibitem[{{Pandey} {et~al.}(2023){Pandey}, {Lehman}, {Baxter}, {Ni}, {Angl{\'e}s-Alc{\'a}zar}, {Genel}, {Villaescusa-Navarro}, {Delgado}, \& {di Matteo}}]{2023MNRAS.525.1779P}
{Pandey}, S., {Lehman}, K., {Baxter}, E.~J., {et~al.} 2023, \mnras, 525, 1779, \dodoi{10.1093/mnras/stad2268}

\bibitem[{{Pandey} {et~al.}(2025{\natexlab{b}}){Pandey}, {Hill}, {Alarcon}, {Alves}, {Amon}, {Anbajagane}, {Andrade-Oliveira}, {Battaglia}, {Baxter}, {Bechtol}, {Becker}, {Bernstein}, {Blazek}, {Bridle}, {Calabrese}, {Camacho}, {Campos}, {Carnero Rosell}, {Carrasco Kind}, {Cawthon}, {Chang}, {Chen}, {Chintalapati}, {Choi}, {Cordero}, {Coulton}, {Crocce}, {Davis}, {DeRose}, {Devlin}, {Diehl}, {Dodelson}, {Doux}, {Drlica-Wagner}, {Eckert}, {Eifler}, {Elvin-Poole}, {Everett}, {Fang}, {Fert{\'e}}, {Fosalba}, {Friedrich}, {Gatti}, {Gaztanaga}, {Giannini}, {Gluscevic}, {Gruen}, {Gruendl}, {Ried Guachalla}, {Harrison}, {Hartley}, {Herner}, {Huang}, {Huff}, {Huterer}, {Jain}, {Jarvis}, {Krause}, {Kuropatkin}, {Kusiak}, {Leget}, {Lemos}, {Liddle}, {Lokken}, {MacCrann}, {McCullough}, {Moodley}, {Muir}, {Myles}, {Navarro-Alsina}, {Omori}, {Park}, {Partridge}, {Porredon}, {Prat}, {Raveri}, {Refregier}, {Rollins}, {Roodman}, {Rosenfeld}, {Ross}, {Rykoff}, {Samuroff}, {Sanchez}, {S{\'a}nchez}, {Secco}, {Sevilla-Noarbe},
  {Shaikh}, {Sheldon}, {Shin}, {Sif{\'o}n}, {To}, {Troja}, {Troxel}, {Tutusaus}, {Varga}, {Weaverdyck}, {Wechsler}, {Wollack}, {Yanny}, {Yin}, {Zhang}, {Zuntz}, {Allam}, {Bacon}, {Bocquet}, {Brooks}, {Burke}, {Carretero}, {Cawthon}, {Costanzi}, {da Costa}, {da Silva Pereira}, {Davis}, {Desai}, {Frieman}, {Garc{\'\i}a-Bellido}, {Gutierrez}, {Hinton}, {Hollowood}, {Honscheid}, {James}, {Jeffrey}, {Lee}, {Marshall}, {Mena-Fern{\'a}ndez}, {Miquel}, {Mohr}, {Ogando}, {Plazas Malag'on}, {Romer}, {Sanchez}, {Santiago}, {Smith}, {Suchyta}, {Swanson}, {Thomas}, {Vikram}, {Walker}, {Weller}, \& {Wiseman}}]{2025arXiv250607432P}
{Pandey}, S., {Hill}, J.~C., {Alarcon}, A., {et~al.} 2025{\natexlab{b}}, arXiv e-prints, arXiv:2506.07432, \dodoi{10.48550/arXiv.2506.07432}

\bibitem[{{Petroff} {et~al.}(2022){Petroff}, {Hessels}, \& {Lorimer}}]{2022A&ARv..30....2P}
{Petroff}, E., {Hessels}, J.~W.~T., \& {Lorimer}, D.~R. 2022, \aapr, 30, 2, \dodoi{10.1007/s00159-022-00139-w}

\bibitem[{{Planck Collaboration} {et~al.}(2020){Planck Collaboration}, {Aghanim}, {Akrami}, {Ashdown}, {Aumont}, {Baccigalupi}, {Ballardini}, {Banday}, {Barreiro}, {Bartolo}, {Basak}, {Battye}, {Benabed}, {Bernard}, {Bersanelli}, {Bielewicz}, {Bock}, {Bond}, {Borrill}, {Bouchet}, {Boulanger}, {Bucher}, {Burigana}, {Butler}, {Calabrese}, {Cardoso}, {Carron}, {Challinor}, {Chiang}, {Chluba}, {Colombo}, {Combet}, {Contreras}, {Crill}, {Cuttaia}, {de Bernardis}, {de Zotti}, {Delabrouille}, {Delouis}, {Di Valentino}, {Diego}, {Dor{\'e}}, {Douspis}, {Ducout}, {Dupac}, {Dusini}, {Efstathiou}, {Elsner}, {En{\ss}lin}, {Eriksen}, {Fantaye}, {Farhang}, {Fergusson}, {Fernandez-Cobos}, {Finelli}, {Forastieri}, {Frailis}, {Fraisse}, {Franceschi}, {Frolov}, {Galeotta}, {Galli}, {Ganga}, {G{\'e}nova-Santos}, {Gerbino}, {Ghosh}, {Gonz{\'a}lez-Nuevo}, {G{\'o}rski}, {Gratton}, {Gruppuso}, {Gudmundsson}, {Hamann}, {Handley}, {Hansen}, {Herranz}, {Hildebrandt}, {Hivon}, {Huang}, {Jaffe}, {Jones}, {Karakci}, {Keih{\"a}nen},
  {Keskitalo}, {Kiiveri}, {Kim}, {Kisner}, {Knox}, {Krachmalnicoff}, {Kunz}, {Kurki-Suonio}, {Lagache}, {Lamarre}, {Lasenby}, {Lattanzi}, {Lawrence}, {Le Jeune}, {Lemos}, {Lesgourgues}, {Levrier}, {Lewis}, {Liguori}, {Lilje}, {Lilley}, {Lindholm}, {L{\'o}pez-Caniego}, {Lubin}, {Ma}, {Mac{\'\i}as-P{\'e}rez}, {Maggio}, {Maino}, {Mandolesi}, {Mangilli}, {Marcos-Caballero}, {Maris}, {Martin}, {Martinelli}, {Mart{\'\i}nez-Gonz{\'a}lez}, {Matarrese}, {Mauri}, {McEwen}, {Meinhold}, {Melchiorri}, {Mennella}, {Migliaccio}, {Millea}, {Mitra}, {Miville-Desch{\^e}nes}, {Molinari}, {Montier}, {Morgante}, {Moss}, {Natoli}, {N{\o}rgaard-Nielsen}, {Pagano}, {Paoletti}, {Partridge}, {Patanchon}, {Peiris}, {Perrotta}, {Pettorino}, {Piacentini}, {Polastri}, {Polenta}, {Puget}, {Rachen}, {Reinecke}, {Remazeilles}, {Renzi}, {Rocha}, {Rosset}, {Roudier}, {Rubi{\~n}o-Mart{\'\i}n}, {Ruiz-Granados}, {Salvati}, {Sandri}, {Savelainen}, {Scott}, {Shellard}, {Sirignano}, {Sirri}, {Spencer}, {Sunyaev}, {Suur-Uski}, {Tauber}, {Tavagnacco},
  {Tenti}, {Toffolatti}, {Tomasi}, {Trombetti}, {Valenziano}, {Valiviita}, {Van Tent}, {Vibert}, {Vielva}, {Villa}, {Vittorio}, {Wandelt}, {Wehus}, {White}, {White}, {Zacchei}, \& {Zonca}}]{2020A&A...641A...6P}
{Planck Collaboration}, {Aghanim}, N., {Akrami}, Y., {et~al.} 2020, \aap, 641, A6, \dodoi{10.1051/0004-6361/201833910}

\bibitem[{{Preston} {et~al.}(2023){Preston}, {Amon}, \& {Efstathiou}}]{2023MNRAS.525.5554P}
{Preston}, C., {Amon}, A., \& {Efstathiou}, G. 2023, \mnras, 525, 5554, \dodoi{10.1093/mnras/stad2573}

\bibitem[{{Prochaska} \& {Zheng}(2019)}]{2019MNRAS.485..648P}
{Prochaska}, J.~X., \& {Zheng}, Y. 2019, \mnras, 485, 648, \dodoi{10.1093/mnras/stz261}

\bibitem[{{Qiu Cheng} {et~al.}(2025){Qiu Cheng}, {Andrew}, {Wang}, \& {Masui}}]{2025arXiv250603258Q}
{Qiu Cheng}, A., {Andrew}, S.~E., {Wang}, H., \& {Masui}, K.~W. 2025, arXiv e-prints, arXiv:2506.03258, \dodoi{10.48550/arXiv.2506.03258}

\bibitem[{{Rafiei-Ravandi} {et~al.}(2020){Rafiei-Ravandi}, {Smith}, \& {Masui}}]{2020PhRvD.102b3528R}
{Rafiei-Ravandi}, M., {Smith}, K.~M., \& {Masui}, K.~W. 2020, \prd, 102, 023528, \dodoi{10.1103/PhysRevD.102.023528}

\bibitem[{{Rafiei-Ravandi} {et~al.}(2021){Rafiei-Ravandi}, {Smith}, {Li}, {Masui}, {Josephy}, {Dobbs}, {Lang}, {Bhardwaj}, {Patel}, {Bandura}, {Berger}, {Boyle}, {Brar}, {Breitman}, {Cassanelli}, {Chawla}, {Adam Dong}, {Fonseca}, {Gaensler}, {Giri}, {Good}, {Halpern}, {Kaczmarek}, {Kaspi}, {Leung}, {Lin}, {Mena-Parra}, {Meyers}, {Michilli}, {M{\"u}nchmeyer}, {Ng}, {Petroff}, {Pleunis}, {Rahman}, {Sanghavi}, {Scholz}, {Shin}, {Stairs}, {Tendulkar}, {Vanderlinde}, \& {Zwaniga}}]{2021ApJ...922...42R}
{Rafiei-Ravandi}, M., {Smith}, K.~M., {Li}, D., {et~al.} 2021, \apj, 922, 42, \dodoi{10.3847/1538-4357/ac1dab}

\bibitem[{{Ravi} {et~al.}(2023){Ravi}, {Catha}, {Chen}, {Connor}, {Cordes}, {Faber}, {Lamb}, {Hallinan}, {Harnach}, {Hellbourg}, {Hobbs}, {Hodge}, {Hodges}, {Law}, {Rasmussen}, {Sharma}, {Sherman}, {Shi}, {Simard}, {Somalwar}, {Squillace}, {Weinreb}, {Woody}, \& {Yadlapalli}}]{2023arXiv230101000R}
{Ravi}, V., {Catha}, M., {Chen}, G., {et~al.} 2023, arXiv e-prints, arXiv:2301.01000, \dodoi{10.48550/arXiv.2301.01000}

\bibitem[{{Reischke} \& {Hagstotz}(2023)}]{2023MNRAS.524.2237R}
{Reischke}, R., \& {Hagstotz}, S. 2023, \mnras, 524, 2237, \dodoi{10.1093/mnras/stad1645}

\bibitem[{Reischke \& {Hagstotz}(2025)}]{2025arXiv250717742R}
Reischke, R., \& {Hagstotz}, S. 2025, arXiv e-prints, arXiv:2507.17742, \dodoi{10.48550/arXiv.2507.17742}

\bibitem[{{Reischke} {et~al.}(2025){Reischke}, {Kova{\v{c}}}, {Nicola}, {Hagstotz}, \& {Schneider}}]{2024arXiv241117682R}
{Reischke}, R., {Kova{\v{c}}}, M., {Nicola}, A., {Hagstotz}, S., \& {Schneider}, A. 2025, The Open Journal of Astrophysics, 8, 13, \dodoi{10.33232/001c.143819}

\bibitem[{{Ried Guachalla} {et~al.}(2025){Ried Guachalla}, {Schaan}, {Hadzhiyska}, {Ferraro}, {Aguilar}, {Ahlen}, {Battaglia}, {Bianchi}, {Bond}, {Brooks}, {Claybaugh}, {Coulton}, {de la Macorra}, {Devlin}, {Dey}, {Doel}, {Dunkley}, {Fanning}, {Forero-Romero}, {Gaztanaga}, {Gontcho}, {Gutierrez}, {Guy}, {Hill}, {Honscheid}, {Juneau}, {Kisner}, {Kremin}, {Lambert}, {Landriau}, {Le Guillou}, {MacCrann}, {Manera}, {Meisner}, {Miquel}, {Moodley}, {Moustakas}, {Mroczkowski}, {Myers}, {Niemack}, {Niz}, {Palanque-Delabrouille}, {Percival}, {P\textbackslash'erez-R\textbackslash`afols}, {Poppett}, {Prada}, {Qu}, {Rossi}, {Sanchez}, {Schlegel}, {Schubnell}, {Seo}, {Sif\textbackslash'on}, {Spergel}, {Sprayberry}, {Tarl\textbackslash'e}, {Vargas-Magana}, {Vavagiakis}, {Weaver}, {Wollack}, \& {Zarrouk}}]{2025arXiv250319870R}
{Ried Guachalla}, B., {Schaan}, E., {Hadzhiyska}, B., {et~al.} 2025, arXiv e-prints, arXiv:2503.19870, \dodoi{10.48550/arXiv.2503.19870}

\bibitem[{{Schneider} {et~al.}(2022){Schneider}, {Giri}, {Amodeo}, \& {Refregier}}]{2022MNRAS.514.3802S}
{Schneider}, A., {Giri}, S.~K., {Amodeo}, S., \& {Refregier}, A. 2022, \mnras, 514, 3802, \dodoi{10.1093/mnras/stac1493}

\bibitem[{{Sharma} {et~al.}(2025){Sharma}, {Krause}, {Ravi}, {Reischke}, {Pranjal R.}, \& {Connor}}]{2025arXiv250418745S}
{Sharma}, K., {Krause}, E., {Ravi}, V., {et~al.} 2025, arXiv e-prints, arXiv:2504.18745, \dodoi{10.48550/arXiv.2504.18745}

\bibitem[{{Shirasaki} {et~al.}(2017){Shirasaki}, {Kashiyama}, \& {Yoshida}}]{2017PhRvD..95h3012S}
{Shirasaki}, M., {Kashiyama}, K., \& {Yoshida}, N. 2017, \prd, 95, 083012, \dodoi{10.1103/PhysRevD.95.083012}

\bibitem[{{Spergel} {et~al.}(2013){Spergel}, {Gehrels}, {Breckinridge}, {Donahue}, {Dressler}, {Gaudi}, {Greene}, {Guyon}, {Hirata}, {Kalirai}, {Kasdin}, {Moos}, {Perlmutter}, {Postman}, {Rauscher}, {Rhodes}, {Wang}, {Weinberg}, {Centrella}, {Traub}, {Baltay}, {Colbert}, {Bennett}, {Kiessling}, {Macintosh}, {Merten}, {Mortonson}, {Penny}, {Rozo}, {Savransky}, {Stapelfeldt}, {Zu}, {Baker}, {Cheng}, {Content}, {Dooley}, {Foote}, {Goullioud}, {Grady}, {Jackson}, {Kruk}, {Levine}, {Melton}, {Peddie}, {Ruffa}, \& {Shaklan}}]{2013arXiv1305.5425S}
{Spergel}, D., {Gehrels}, N., {Breckinridge}, J., {et~al.} 2013, arXiv e-prints, arXiv:1305.5425, \dodoi{10.48550/arXiv.1305.5425}

\bibitem[{{The LSST Dark Energy Science Collaboration} {et~al.}(2018){The LSST Dark Energy Science Collaboration}, {Mandelbaum}, {Eifler}, {Hlo{\v{z}}ek}, {Collett}, {Gawiser}, {Scolnic}, {Alonso}, {Awan}, {Biswas}, {Blazek}, {Burchat}, {Chisari}, {Dell'Antonio}, {Digel}, {Frieman}, {Goldstein}, {Hook}, {Ivezi{\'c}}, {Kahn}, {Kamath}, {Kirkby}, {Kitching}, {Krause}, {Leget}, {Marshall}, {Meyers}, {Miyatake}, {Newman}, {Nichol}, {Rykoff}, {Sanchez}, {Slosar}, {Sullivan}, \& {Troxel}}]{2018arXiv180901669T}
{The LSST Dark Energy Science Collaboration}, {Mandelbaum}, R., {Eifler}, T., {et~al.} 2018, arXiv e-prints, arXiv:1809.01669, \dodoi{10.48550/arXiv.1809.01669}

\bibitem[{{The RAIL Team} {et~al.}(2025){The RAIL Team}, {van den Busch}, {Charles}, {Cohen-Tanugi}, {Crafford}, {Crenshaw}, {Dagoret}, {De-Santiago}, {De Vicente}, {Hang}, {Joachimi}, {Joudaki}, {Bryce Kalmbach}, {Kannawadi}, {Liang}, {Lynn}, {Malz}, {Mandelbaum}, {Merz}, {Moskowitz}, {Oldag}, {Ruiz-Zapatero}, {Rahman}, {Rau}, {Schmidt}, {Scora}, {Shirley}, {St{\"o}lzner}, {Toribio San Cipriano}, {Tortorelli}, {Yan}, {Zhang}, \& {the Dark Energy Science Collaboration}}]{2025arXiv250502928T}
{The RAIL Team}, {van den Busch}, J.~L., {Charles}, E., {et~al.} 2025, arXiv e-prints, arXiv:2505.02928, \dodoi{10.48550/arXiv.2505.02928}

\bibitem[{{Thornton} {et~al.}(2013){Thornton}, {Stappers}, {Bailes}, {Barsdell}, {Bates}, {Bhat}, {Burgay}, {Burke-Spolaor}, {Champion}, {Coster}, {D'Amico}, {Jameson}, {Johnston}, {Keith}, {Kramer}, {Levin}, {Milia}, {Ng}, {Possenti}, \& {van Straten}}]{2013Sci...341...53T}
{Thornton}, D., {Stappers}, B., {Bailes}, M., {et~al.} 2013, Science, 341, 53, \dodoi{10.1126/science.1236789}

\bibitem[{{To} {et~al.}(2024){To}, {Pandey}, {Krause}, {Dalal}, {Anbajagane}, \& {Weinberg}}]{2024JCAP...07..037T}
{To}, C.-H., {Pandey}, S., {Krause}, E., {et~al.} 2024, \jcap, 2024, 037, \dodoi{10.1088/1475-7516/2024/07/037}

\bibitem[{{Tr{\"o}ster} {et~al.}(2022){Tr{\"o}ster}, {Mead}, {Heymans}, {Yan}, {Alonso}, {Asgari}, {Bilicki}, {Dvornik}, {Hildebrandt}, {Joachimi}, {Kannawadi}, {Kuijken}, {Schneider}, {Shan}, {van Waerbeke}, \& {Wright}}]{2022A&A...660A..27T}
{Tr{\"o}ster}, T., {Mead}, A.~J., {Heymans}, C., {et~al.} 2022, \aap, 660, A27, \dodoi{10.1051/0004-6361/202142197}

\bibitem[{{Vanderlinde} {et~al.}(2019){Vanderlinde}, {Liu}, {Gaensler}, {Bond}, {Hinshaw}, {Ng}, {Chiang}, {Stairs}, {Brown}, {Sievers}, {Mena}, {Smith}, {Bandura}, {Masui}, {Spekkens}, {Belostotski}, {Dobbs}, {Turok}, {Boyle}, {Rupen}, {Landecker}, {Pen}, \& {Kaspi}}]{2019clrp.2020...28V}
{Vanderlinde}, K., {Liu}, A., {Gaensler}, B., {et~al.} 2019, in Canadian Long Range Plan for Astronomy and Astrophysics White Papers, Vol. 2020, 28, \dodoi{10.5281/zenodo.3765414}

\bibitem[{{Wang} {et~al.}(2025){Wang}, {Masui}, {Andrew}, {Fonseca}, {Gaensler}, {Joseph}, {Kaspi}, {Kharel}, {Lanman}, {Leung}, {Mas-Ribas}, {Mena-Parra}, {Nimmo}, {Pearlman}, {Pen}, {Prochaska}, {Raikman}, {Shin}, {Siegel}, {Smith}, \& {Stairs}}]{2025arXiv250608932W}
{Wang}, H., {Masui}, K., {Andrew}, S., {et~al.} 2025, arXiv e-prints, arXiv:2506.08932, \dodoi{10.48550/arXiv.2506.08932}

\bibitem[{{Yao} {et~al.}(2023){Yao}, {Shan}, {Zhang}, {Jullo}, {Kneib}, {Yu}, {Zu}, {Brooks}, {de la Macorra}, {Doel}, {Font-Ribera}, {Gontcho}, {Kisner}, {Landriau}, {Meisner}, {Miquel}, {Nie}, {Poppett}, {Prada}, {Schubnell}, {Vargas Magana}, \& {Zhou}}]{2023MNRAS.524.6071Y}
{Yao}, J., {Shan}, H., {Zhang}, P., {et~al.} 2023, \mnras, 524, 6071, \dodoi{10.1093/mnras/stad2221}

\bibitem[{{Yuan} {et~al.}(2024){Yuan}, {Zhang}, {Ross}, {Donald-McCann}, {Hadzhiyska}, {Wechsler}, {Zheng}, {Alam}, {Gonzalez-Perez}, {Aguilar}, {Ahlen}, {Bianchi}, {Brooks}, {de la Macorra}, {Fanning}, {Forero-Romero}, {Honscheid}, {Ishak}, {Kehoe}, {Lasker}, {Landriau}, {Manera}, {Martini}, {Meisner}, {Miquel}, {Moustakas}, {Nadathur}, {Newman}, {Nie}, {Percival}, {Poppett}, {Rocher}, {Rossi}, {Sanchez}, {Samushia}, {Schubnell}, {Seo}, {Tarl{\'e}}, {Weaver}, {Yu}, {Zhou}, \& {Zou}}]{2024MNRAS.530..947Y}
{Yuan}, S., {Zhang}, H., {Ross}, A.~J., {et~al.} 2024, \mnras, 530, 947, \dodoi{10.1093/mnras/stae359}

\bibitem[{{Zhang} {et~al.}(2024){Zhang}, {Comparat}, {Ponti}, {Merloni}, {Nandra}, {Haberl}, {Locatelli}, {Zhang}, {Sanders}, {Zheng}, {Liu}, {Popesso}, {Liu}, {Truong}, {Pillepich}, {Predehl}, {Salvato}, {Shreeram}, {Yeung}, \& {Ni}}]{2024A&A...690A.267Z}
{Zhang}, Y., {Comparat}, J., {Ponti}, G., {et~al.} 2024, \aap, 690, A267, \dodoi{10.1051/0004-6361/202449412}

\bibitem[{{Zheng} {et~al.}(2007){Zheng}, {Coil}, \& {Zehavi}}]{2007ApJ...667..760Z}
{Zheng}, Z., {Coil}, A.~L., \& {Zehavi}, I. 2007, \apj, 667, 760, \dodoi{10.1086/521074}

\bibitem[{{Zheng} {et~al.}(2005){Zheng}, {Berlind}, {Weinberg}, {Benson}, {Baugh}, {Cole}, {Dav{\'e}}, {Frenk}, {Katz}, \& {Lacey}}]{2005ApJ...633..791Z}
{Zheng}, Z., {Berlind}, A.~A., {Weinberg}, D.~H., {et~al.} 2005, \apj, 633, 791, \dodoi{10.1086/466510}

\bibitem[{{Zhou} {et~al.}(2023){Zhou}, {Dey}, {Newman}, {Eisenstein}, {Dawson}, {Bailey}, {Berti}, {Guy}, {Lan}, {Zou}, {Aguilar}, {Ahlen}, {Alam}, {Brooks}, {de la Macorra}, {Dey}, {Dhungana}, {Fanning}, {Font-Ribera}, {Gontcho}, {Honscheid}, {Ishak}, {Kisner}, {Kov{\'a}cs}, {Kremin}, {Landriau}, {Levi}, {Magneville}, {Manera}, {Martini}, {Meisner}, {Miquel}, {Moustakas}, {Myers}, {Nie}, {Palanque-Delabrouille}, {Percival}, {Poppett}, {Prada}, {Raichoor}, {Ross}, {Schlafly}, {Schlegel}, {Schubnell}, {Tarl{\'e}}, {Weaver}, {Wechsler}, {Y{\'e}che}, \& {Zhou}}]{2023AJ....165...58Z}
{Zhou}, R., {Dey}, B., {Newman}, J.~A., {et~al.} 2023, \aj, 165, 58, \dodoi{10.3847/1538-3881/aca5fb}

\end{thebibliography}
\bibliographystyle{aasjournal}

\end{document}